\documentclass[journal,10pt]{IEEEtran}

\usepackage{flushend}
\usepackage{csquotes}
\usepackage{graphicx}
\usepackage{amsmath}
\usepackage{amsthm}
\usepackage{amssymb}
\usepackage{color}
\usepackage{algorithm}
\usepackage{algpseudocode}
\usepackage[normalem]{ulem}
\usepackage{cite}
\useunder{\uline}{\ul}{}
\usepackage{url}
\usepackage{setspace}
\usepackage{comment}
\usepackage{booktabs}
\usepackage{hyperref}

\usepackage{tabularx}
\usepackage{multirow}

\DeclareMathOperator*{\minimize}{minimize}

\ifCLASSOPTIONcompsoc
\usepackage[caption=false,font=normalsize,labelfont=sf,textfont=sf]{subfig}
\else
\usepackage[caption=false,font=footnotesize]{subfig}
\fi

\usepackage{cuted,tcolorbox,lipsum}
\definecolor{infocolor}{RGB}{245,245,245}
\usepackage{multicol}
\usepackage{capt-of}

\theoremstyle{definition}

{
	\theoremstyle{plain}
	
}

\title{Unfolding-Aided Bootstrapped Phase Retrieval in Optical Imaging \vspace{-0.2em}}

\author{\IEEEauthorblockN{ 
		Samuel Pinilla$^{1}$, Kumar Vijay Mishra$^{3}$, Igor Shevkunov$^{2}$, Mojtaba Soltanalian$^{4}$,\\ Vladimir Katkovnik$^{2,5}$, and Karen Egiazarian$^{2,5}$ \\
		\textit{$^{1}$University of Manchester at Harwell Science and Innovation campus, Oxfordshire, UK} \\
		\textit{$^{2}$ Faculty of Information Technology and Communication Sciences, Tampere University, 33720, Finland} \\
		\textit{$^{3}$United States CCDC Army Research Laboratory, Adelphi, MD 20783, USA} \\
		\textit{$^{4}$ University of Illinois Chicago, Chicago, IL 60607, USA}\\
		\textit{$^{5}$Noiseless Imaging Ltd., Tampere, 33720, Finland}\vspace{-0.8em}}\vspace{-20pt}
}

\begin{document}
	
	\maketitle
	
	\IEEEpeerreviewmaketitle
	\begin{abstract}
		Phase retrieval in optical imaging refers to the recovery of a complex signal from phaseless data acquired in the form of its diffraction patterns. These patterns are acquired through a system with a coherent light source that employs a diffractive optical element (DOE) to modulate the scene resulting in \textit{coded diffraction patterns} at the sensor. Recently, the hybrid approach of model-driven network or deep \textit{unfolding} has emerged as an effective alternative to conventional model-based and learning-based phase retrieval techniques because it allows for bounding the complexity of algorithms while also retaining their efficacy. Additionally, such hybrid approaches have shown promise in improving the design of DOEs that follow theoretical uniqueness conditions. There are opportunities to exploit novel experimental setups and resolve even more complex DOE phase retrieval applications. This paper presents an overview of algorithms and applications of deep unfolding for \textit{bootstrapped} - regardless of near, middle, and far zones - phase retrieval.
	\end{abstract}
	
	\section{Introduction}
	Phase retrieval in diffractive optical imaging (DOI) deals with the estimation of a complex domain signal from magnitude-only data in the form of its diffraction patterns \cite{pinilla2018phase}. This problem has been at the center of exciting developments in DOI~\cite{ajerez,guerrero,pinilla2018codedXC} during the last decade to improve achievable spatial resolution \cite{jbacca,kocsis2021ssr,katkovnik2017computational}, lighter optical setups \cite{kocsis2021ssr}, and better detection \cite{ajerez}. When the diffraction patterns (phaseless data) are acquired through a setup that employs coherent light and a diffractive optical element (DOE) (also as coded aperture) to modulate the scene, phase \textit{coded diffraction patterns} (CDP) are obtained. Typically this setup allows to acquire several snapshots of the scene by changing the spatial configuration of the DOE. The inclusion of DOE in the phase retrieval setting paves the way for uniqueness guarantees for all two-dimensional (2-D) signals up to a unimodular constant \cite{guerrero,candes2015phase,gross2017improved}, overcoming the classical result in \cite{hayes1982reconstruction} that holds for only non-reducible signals.
	
	The CDP is experimentally acquired in three diffraction zones: near, middle, and far~\cite{poon2014introduction,guerrero}. Mathematically, assuming a diagonal matrix $\boldsymbol{D}_{\ell}\in \mathbb{C}^{n\times n}$ modelling the DOE for the $\ell$-th snapshot, $\ell=1,\dots,L$, the CDP consists of quadratic equations of the form $y_{k,i,\ell}=\lvert \boldsymbol{a}_{k,i}^{H}\boldsymbol{D}_{\ell}\boldsymbol{x}\rvert$, where $\boldsymbol{a}_{k,i}\in\mathbb{C}^{n}$ are the known wavefront propagation vectors associated with the $k$-th diffraction for $i=1,\cdots,n$, $\boldsymbol{x}\in\mathbb{C}^{n}$ is the unknown scene of interest, and $k=1, 2, 3$ indexes the near, middle, and far zones, respectively. By harnessing specific properties of each diffraction zone, several advances in imaging applications have been made. The near zone is exploited in scanning near-field optical microscopy~\cite{durig1986near}, Raman imaging~\cite{jahncke1995raman}, and spectroscopy~\cite{hess1994near}. Applications such as Fresnel holography~\cite{poon2014introduction} and lensless imaging~\cite{shimano2018lensless} take advantage of the middle diffraction zone to develop new acquisition imaging devices~\cite{shimano2018lensless}, and optical elements such as Fresnel lenses~\cite{sao2018lensless}. The far or Fraunhofer diffraction zone has been instrumental in the development of applications such as crystallography, astronomical imaging, and microscopy~\cite{pinilla2020single}. 
	
	Motivated by the above mentioned applications and guided by the analytical models of the physical setups for a given diffraction zone, \cite{guerrero} has showed that the DOE and the phase retrieval algorithm need to be jointly designed to guarantee unique recovery -- \textit{these theoretical results were not possible a decade ago}. The entries of each matrix $\boldsymbol{D}_{\ell}$ (DOE for $\ell$-th snapshot) were assumed to be independent identically distributed (i.i.d) from a discrete random variable $d$ obeying $\lvert d\rvert \leq 1$. Interestingly, \cite{guerrero} also shows that the \textit{design criteria for DOEs is independent of the diffraction zone.} This result provides a \textit{bootstrapped} (regardless of the diffraction zone) phase retrieval in optical imaging. Table~\ref{tab:1} summarizes common DOE phase retrieval applications.
	
	The role of DOEs in recovering the phase across all diffraction zones has been studied in~\cite{katkovnik2017computational,ajerez,arguello2021shift,monakhova2019learned,karitans2019optical,tong2021quantitative}, where either the DOE was modelled by specific measurement setups or machine learning (ML) was deployed to obtain data-driven model-free recovery. While very flexible and powerful, model-based methods also yield significant reconstruction artefacts arising from imperfect system models, mismatch, calibration errors, hand-tuned parameters, and hand-picked regularizers. Furthermore, these methods take hundreds to thousands of iterations to converge, which is unacceptable for real-time imaging. These shortcomings are overcome by employing deep neural networks, whose expressive power allows training such that the resulting network acts as an estimator of the true signal. For example, \cite{morales2021object} proposed a deep neural network for object classification from CDP achieving high stability in the classification performance when all CDPs are not employed. However, neural networks are treated as black boxes leading to limited interpretability and, therefore, inadequate control. Further, the training times could be prohibitively longer.
	
	In this context, a hybrid model-based and data-driven technique has the potential to further improve data acquisition by effectively mitigating the adverse effects of both techniques \cite{wang2020phase, wang2021phase}. This so-called deep \textit{unfolding} technique effectively facilitates the design of model-aware deep architectures based on well-established iterative signal processing techniques~\cite{naimipour2020upr,gregor2010learning,ongie2020deep}. Unfolding exploits the data to enhance accuracy and performance so that the resulting interpretability leads to trusted outcomes, requiring lesser data, and faster convergence rate. Another related work in \cite{gregor2010learning} viewed the model-based Iterative Shrinkage Thresholding Algorithm (ISTA) as a recurrent neural network, where its activation functions are the shrinkage operator. In the context of inverse problems in imaging, \cite{ongie2020deep} provides further insights into the theory and implementations of unrolling. The related terms include \textit{unrolling} \cite{liu2019deep} and \textit{unrectifying} \cite{hwang2019rectifying}. Whereas the former is used for the algorithms employed by unfolding networks (or simply synonymously with unfolding), the latter refers to the process of transforming nonlinear activations in neural networks into data-dependent linear equations and constraints. Very recently, unfolded networks have been proposed for both conventional and sparse phase retrieval problems \cite{naimipour2020unfolded}. In this paper, we present a \textit{bootstrapped} overview of deep unfolding methods for imaging tasks in the CDP phase retrieval setting, to improve the design of DOEs that can follow theoretical conditions to meet uniqueness. We also summarize the use of designed DOEs together with their unfolded phase retrieval (UPR) algorithms in terms of robustness against noise \cite{morales2022learning}, with/without priors on the scene \cite{naimipour2020unfolded}, and smoothness \cite{estupinan2021deep}. Finally, we review progress in application areas that enable experimental setups to implement these techniques to address even more ill-posed problems.\vspace{-0.5em}
	
	\begin{strip}
		\vspace{-1em}
		\begin{tcolorbox}[colback=infocolor,title={}]
			\vspace{-2em}
			\begin{table}[H]
				\centering
				\caption{Common DOE-Based Optical Imaging Applications}
				\begin{tabular}{p{7cm} p{9.5cm}}
					\hline
					\textbf{Application and Setup} & \textbf{Description}\\
					\hline 
					\begin{minipage}{0.4\columnwidth}
						\centering
						\includegraphics[width=0.8\linewidth]{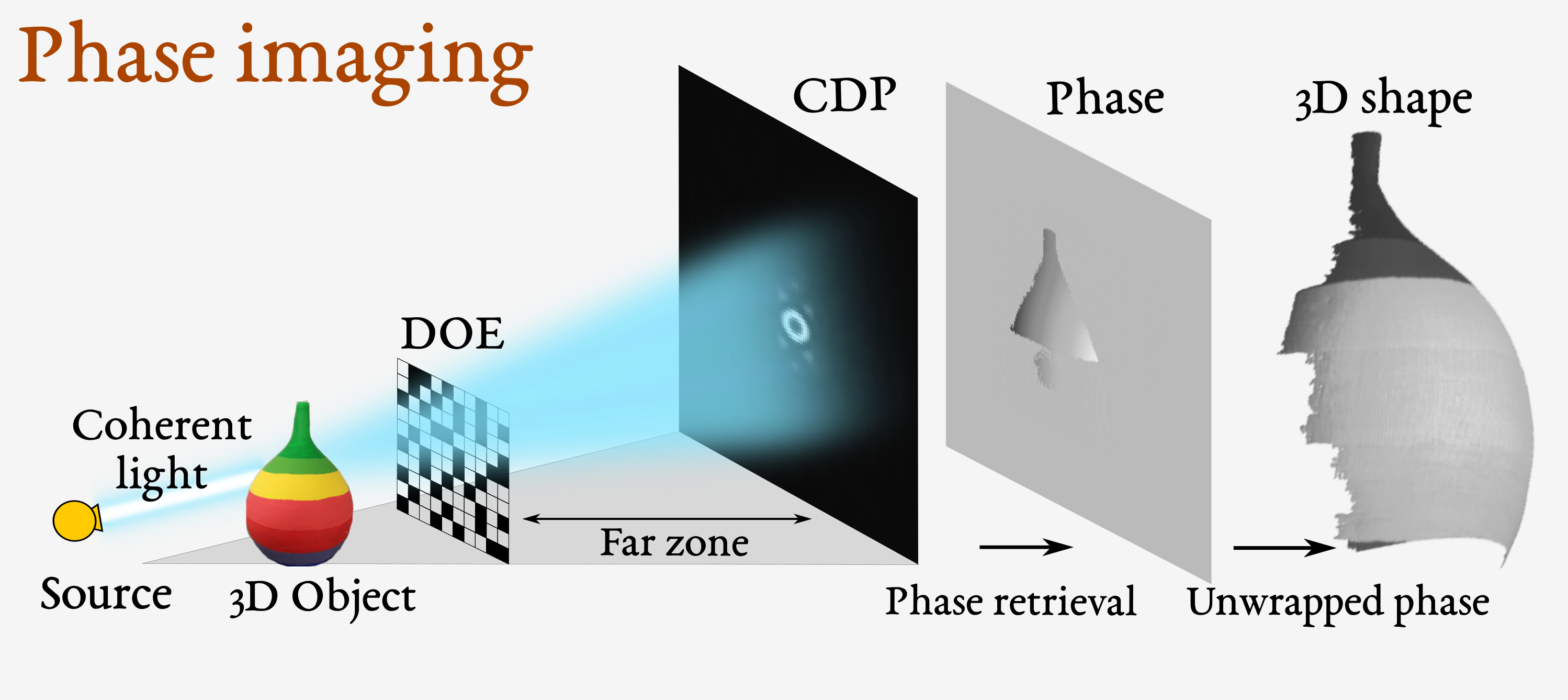}
					\end{minipage} & \begin{minipage}{0.57\columnwidth}
						\vspace{0.3em} Reconstruction of three-dimensional (3-D) shape of an object via phase retrieval. Far zone scenario consists of estimating the optical phase of the object by low-pass-filtering the leading eigenvector of a carefully constructed matrix \cite{pinilla2020single}. 
						\vspace{0.3em}
					\end{minipage}\hspace{0.1em}\\
					\hline\begin{minipage}{0.4\columnwidth}
						\centering
						\includegraphics[width=0.8\linewidth]{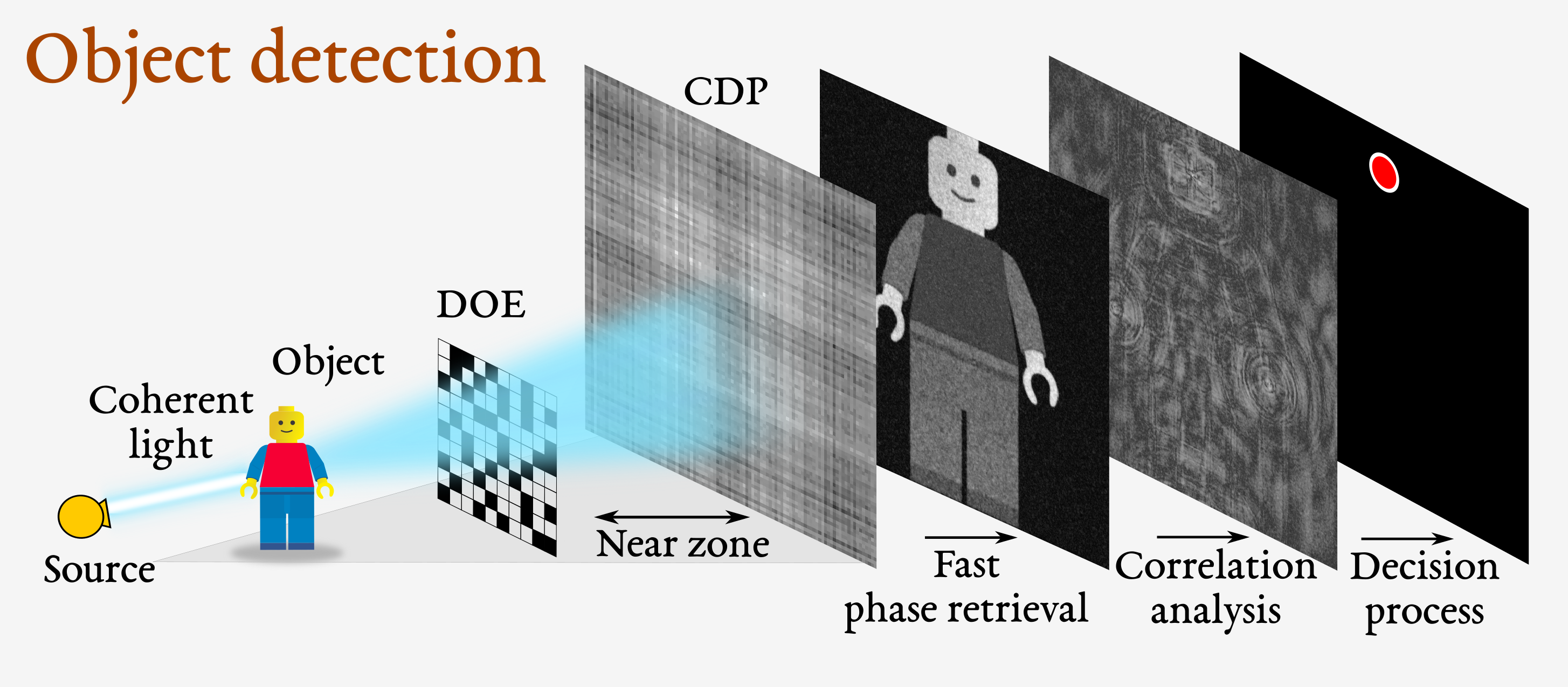}
					\end{minipage} & \begin{minipage}{0.57\columnwidth}
						\vspace{0.3em}Optical phase used to detect objects within a scene. Near zone CDP for rapid detection using cross-correlation analysis to detect the target using its optical phase as a discriminant \cite{ajerez}. \vspace{0.3em}
					\end{minipage}\hspace{0.1em}\\
					\hline\begin{minipage}{0.42\columnwidth}
						\centering
						\includegraphics[width=0.8\linewidth]{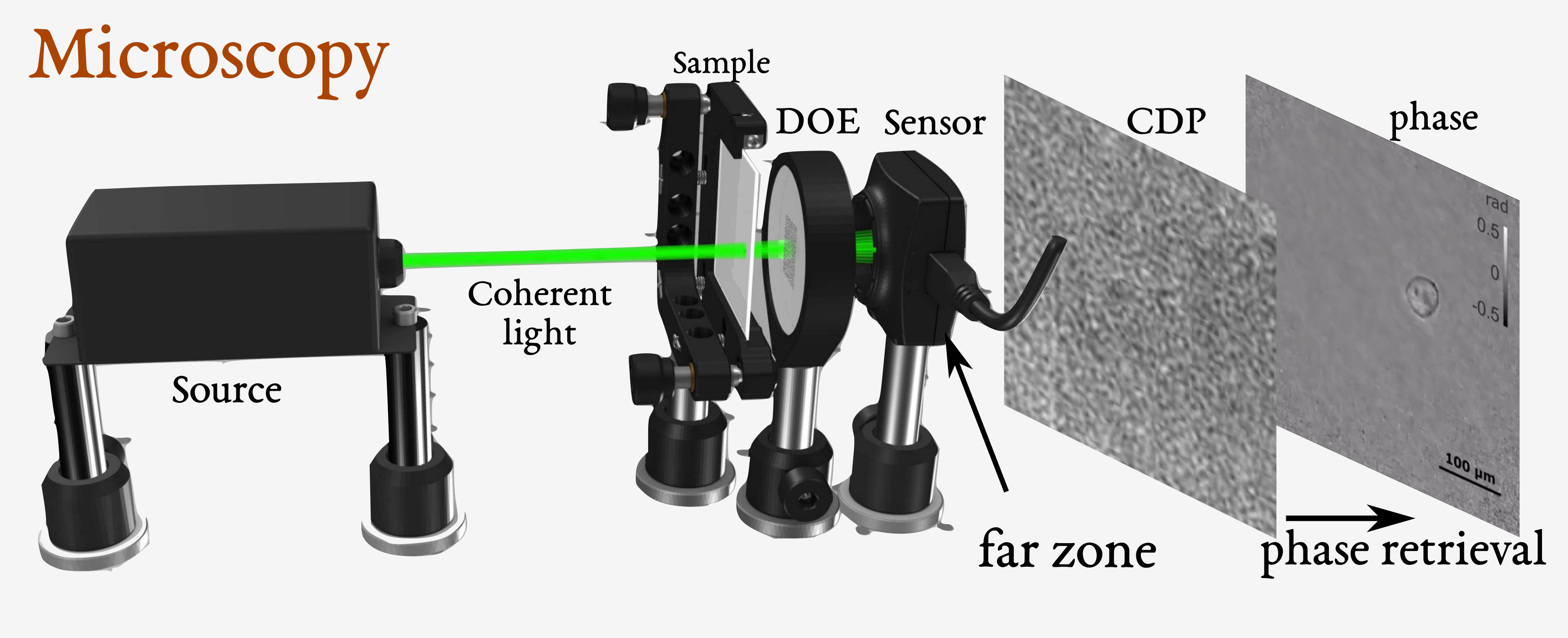}
					\end{minipage} & \begin{minipage}{0.57\columnwidth}
						\vspace{0.3em} Reconstruction of the object wavefront. Lensless single-shot phase retrieval for pixel super-resolution phase imaging in the middle zone \cite{kocsis2021ssr}. Noise is suppressed by a combination of sparse- and deep learning-based filters. Single-shot allows recording of dynamic scenes (frame rate limited only by the camera).\vspace{0.3em}
					\end{minipage}\hspace{0.1em}\\
					\hline
					\begin{minipage}{0.42\columnwidth}
						\centering
						\includegraphics[width=0.8\linewidth]{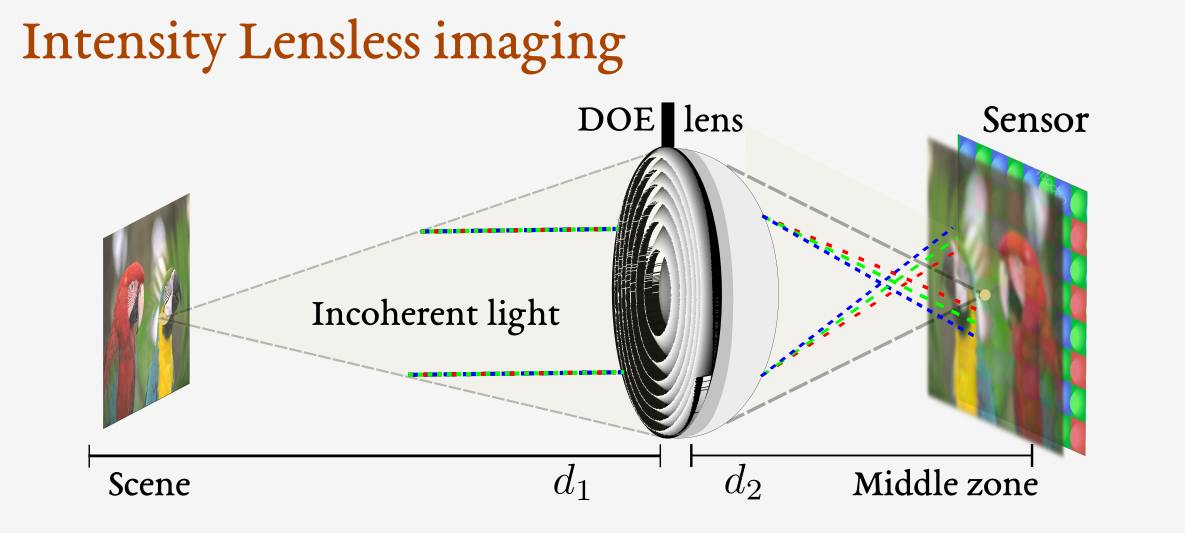}
					\end{minipage} & \begin{minipage}{0.57\columnwidth}
						\vspace{0.3em} Computational imaging with DOEs. Phase retrieval appears in the design of the DOEs instead of the wavefront propagation model. Effectively designed DOE for all-in-focus intensity imaging in the middle zone \cite{rostami2021power}.\vspace{0.3em}
					\end{minipage}\hspace{0.1em}\\
					\hline
				\end{tabular}
				\label{tab:1}
			\end{table}
		\end{tcolorbox}
		\vspace{-1em}
	\end{strip} 
	
	\section{Optimal DOE Design}\vspace{-0.5em}
	The design principles \cite{guerrero} of DOEs require that the set of matrices $\boldsymbol{D}_{\ell}$ satisfies $\mathbb{E}\left[\sum_{\ell=1}^{L}\overline{\boldsymbol{D}}_{\ell}\boldsymbol{D}_{\ell}\right]=r\boldsymbol{I}$, $\ell=1,\dots,L$, to guarantee uniqueness in recovery for all 2-D scenes with high probability (regardless of the zone), with $0<r\leq L$, where $L\geq c_{0}n$ for some large constant $c_{0}>0$, with $\boldsymbol{I}$ as the identity matrix, and $\mathbb{E}[\cdot]$ is the statistical expectation. The implication is that the \textit{variance of the entries} of matrices $\boldsymbol{D}_{\ell}$ needs to be designed along all snapshots to guarantee unique recovery. More information is collected from the scene when this variance is maximized along the snapshots \cite{guerrero,pinilla2018coded}. In the following, we summarize the design principles for two common scenarios: without any prior assumption on the scene and when the scene is known to be sparse in some basis. As explained later, unfolding is applicable to both scenarios. \vspace{-0.5em}
	
	\subsection{Design without priors on the scene}
	This is accomplished in either by targeting the design of the DOEs in the image registration process or obtaining the sensing matrix $\boldsymbol{A}_{k}$ to acquire CDP at the $k$-th diffraction zone \textit{co-designed} jointly with phase retrieval algorithm. It is the latter where unfolding is commonly employed (see box).
	\subsubsection{Greedy DOE methodology}
	For ease of exposition, consider $\boldsymbol{C}^{\ell}\in \mathbb{C}^{N\times N}$ to be the 2-D version of the diagonal matrix $\boldsymbol{D}_{\ell}$ whose entries are $(\boldsymbol{D}_{\ell})_{q,q}=(\boldsymbol{C}^{\ell})_{q-vN,v+1}$, for $v=\lfloor \frac{q-1}{N} \rfloor$, $q=1,\cdots,n$, and $n=N^{2}$. The design principles are based on
	considering an admissible random variable $d=\{e_{1},e_{2},e_{3},e_{4}\}$ with probability $\{\frac{1}{4},\frac{1}{4},\frac{1}{4},\frac{1}{4}\}$, respectively, assuming that $L=C_{d}b$ for some integer $b>0$, where $C_{d}=4$ is the number of coding elements. 
	
	In particular, DOE design takes temporal correlation and spatial separation into account. For temporal correlation, the condition on the 2-D version $\boldsymbol{C}^{\ell}$ of $\boldsymbol{D}_{\ell}$ can be satisfied if matrices $\boldsymbol{C}^{\ell}$ are constrained to have all the coding elements of $d$ along the $L$-projections at each particular spatial position of the ensemble \cite{pinilla2018coded} to directly maximize the variance of the entries along snapshots. It means that $ \{ (\boldsymbol{C}^{\ell})_{s,u} | \ell=1,\cdots,L \}$ contains $b$ times each possible value of $d$ for any $s,u\in \{1,\cdots,N\}$. For spatial separation, it has been shown that a set of DOEs with an equi-spaced distribution of the coding elements improves the ability of data $y_{k,i,\ell}$ to uniquely identify the scene $\boldsymbol{x}$ and increases the image reconstruction quality~\cite{correa2016spatiotemporal,pinilla2018coded}. 
	
	\begin{strip}
		\vspace{-1.5em}
		\begin{tcolorbox}[colback=infocolor,title={Comparison between DOE design methodologies}]
			\begin{multicols}{2}
				\begin{minipage}{1.0\columnwidth}		
					Greedy methods design the DOE itself (instead of the sensing matrix). The experimental performance of this strategy for phase retrieval typically not as expected because of imperfect system modelling, reconstruction artefacts such as mismatch, calibration errors, hand-tuned parameters, and even poorer performance for single snapshot.
					\begin{center}
						\includegraphics[width=0.8\linewidth]{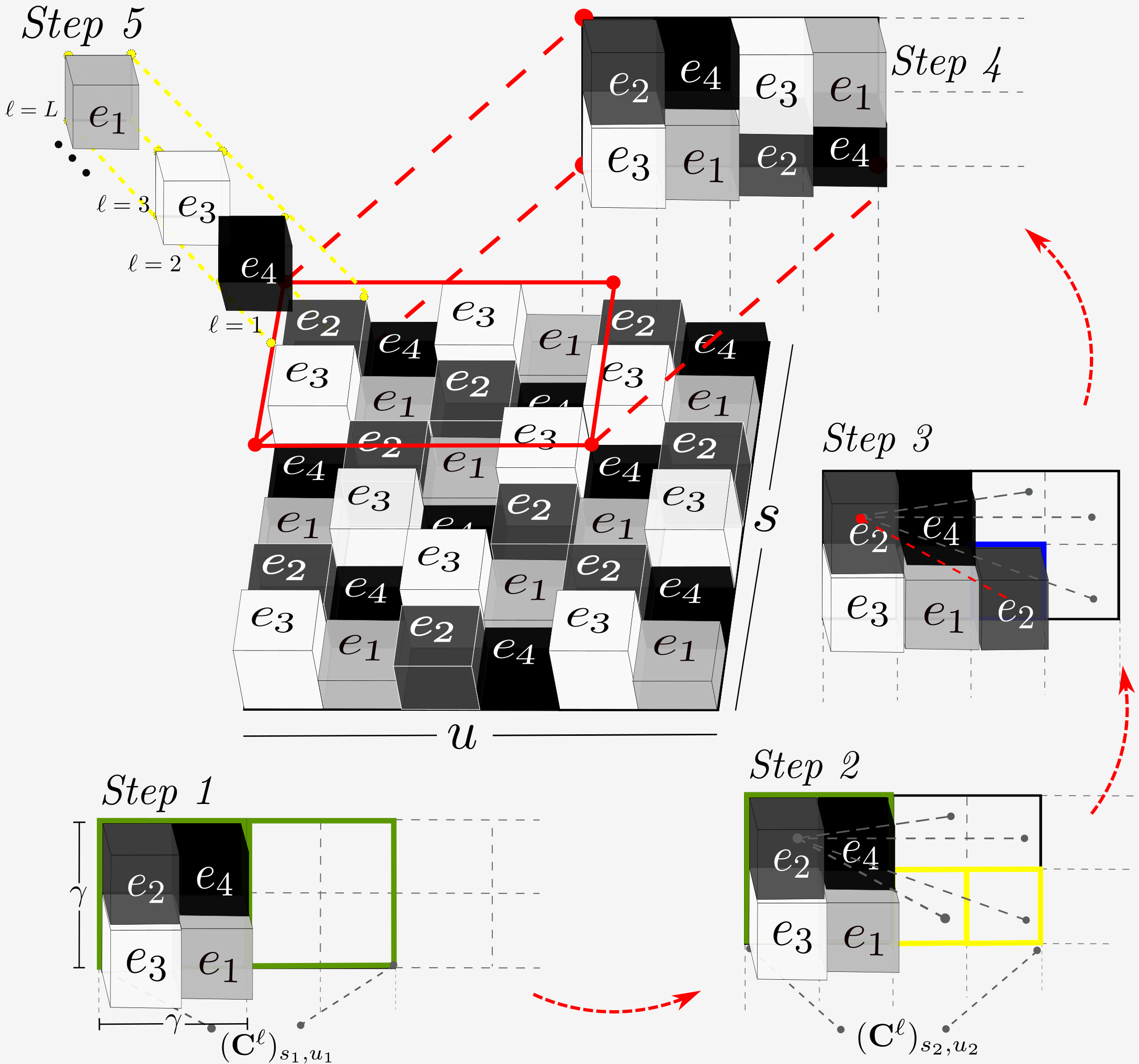}
						\captionof{figure}{DOE design strategy using admissible random variable $d=\{e_1,e_2,e_3,e_4\}$ with probability $\{\frac{1}{4},\frac{1}{4},\frac{1}{4},\frac{1}{4}\}$, respectively and $C_d$~$=4$.}
						\label{fig:codes}
						\vspace{0.5em}
					\end{center}
				\end{minipage}
				
				\begin{minipage}{1.0\columnwidth}
					In unfolding-based techniques, the sensing matrix is designed across all diffraction zones. The performance of the unfolding design method overcomes the greedy methodology because the sensing matrix and phase retrieval algorithm are jointly designed; and the reconstruction algorithm is adjusted for the single snapshot extreme case.
					\begin{center}
						\includegraphics[width=0.55\linewidth]{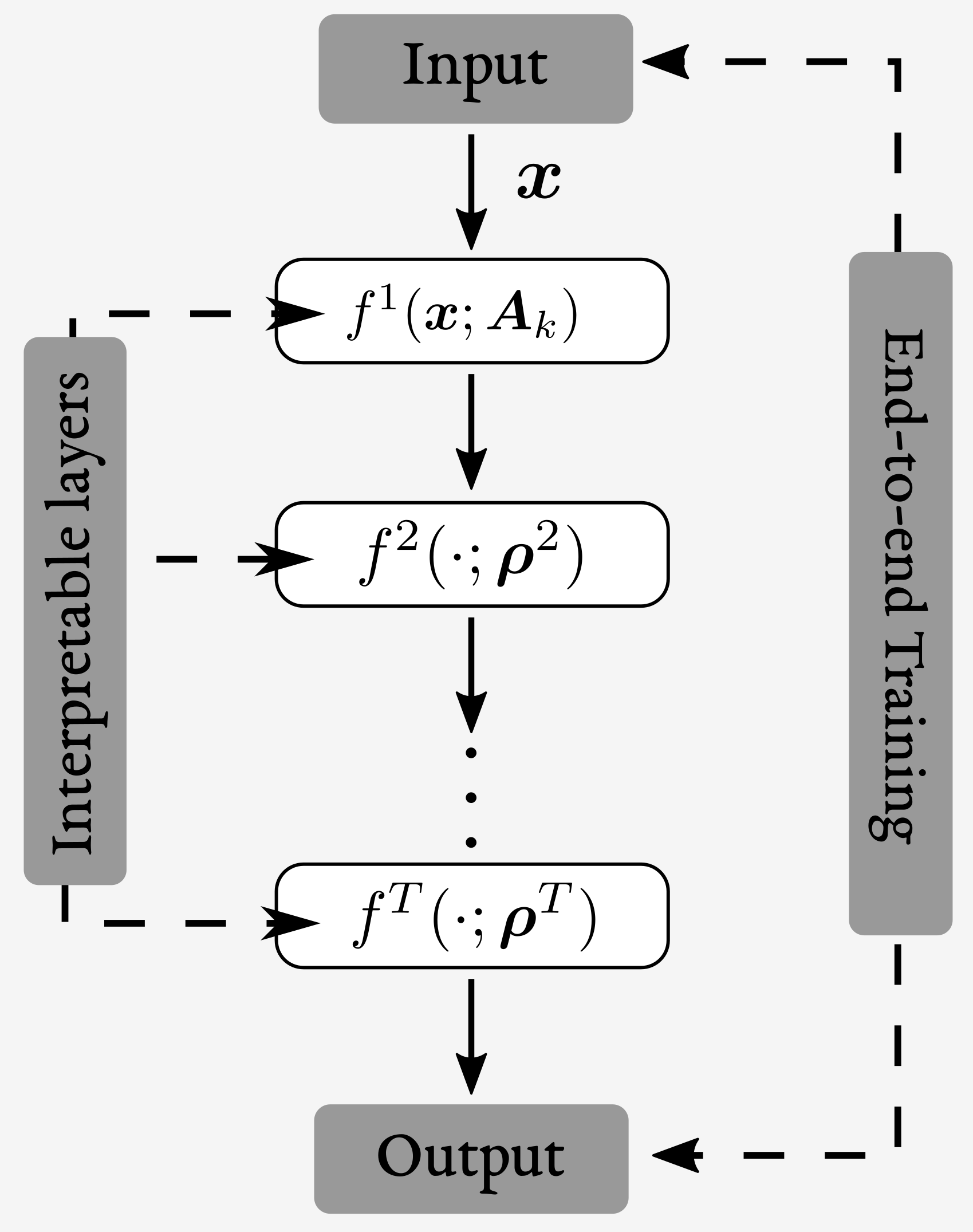}
						\vspace{-0.5em}
						\captionof{figure}{High-level overview of unfolding method resulting in the network layers $f^{1},\dots,f^{T}$. The first layer $f^{1}(\boldsymbol{x};\boldsymbol{A}_{k})=\lvert \boldsymbol{A}_{k}\boldsymbol{x} \rvert$ to model the CDP for the $k$-th diffraction zone. Each iteration for phase retrieval is transferred into network parameters $\boldsymbol{\rho}^{2},\dots,\boldsymbol{\rho}^{T}$. These parameters are learned from training data sets through end-to-end training.}
						\label{fig:basicNetwork}
					\end{center}
				\end{minipage}
			\end{multicols} 
		\end{tcolorbox}
		\vspace{-1em}
	\end{strip}
	
	The overall design strategy comprises the following steps.\\
	\textit{Step 1: Split into several cells. }The matrix $\boldsymbol{C}^{\ell}$ is divided into squared cells (green squares highlighted Figure \ref{fig:codes}). The positions of all $e_{i}$ of $d$ in the first cell of $\boldsymbol{C}^{\ell}$ are chosen uniformly at random as $(\boldsymbol{C}^{\ell})_{s,u}\sim \mathcal{U}[\{e_{1},\cdots,e_{4}\}-\{(\boldsymbol{C}^{1})_{s,u},\cdots,(\boldsymbol{C}^{\ell-1})_{s,u} \}]$ for $s,u$ in the cell. Note that set subtraction $\{e_{1},\cdots,e_{4}\}-\{(\boldsymbol{C}^{1})_{s,u},\cdots,(\boldsymbol{C}^{\ell-1})_{s,u} \}$ guarantees that a different coding element of $d$ is chosen for each of the first $\ell-1$ coded apertures.
	\\
	\textit{Step 2: Determine coding element location. }Move to the cell on the right to determine at random a proper position for each coding element. These positions are determined by maximizing the Manhattan distance between pixels $(\boldsymbol{C}^{\ell})_{s_1,u_1}$ and $(\boldsymbol{C}^{\ell})_{s_2,u_2}$ in the cell on the right, respectively. The positions maximizing this distance for the $e_{2}$ element is one of the two highlighted yellow squares. 
	\\
	\textit{Step 3: Maximal spatial separation. } Define $\Omega_{e_{i}}$ be the set of distances between the pixel with value $e_{i}$ in the first cell and the positions $(s_{2},u_{2})$ of the next right cell. The positions $\mathcal{R}_{e_{i}}$ of maximal distance (as in previous step) are obtained as $\mathcal{R}_{e_{i}} = \operatorname{arg\,max}\Omega_{e_{i}} $. Define the set $\mathcal{B}_{e_{i}}=\{(s,u)\in \mathcal{R}_{e_{i}}| (\boldsymbol{C}^{k})_{s,u}=e_{i}, \text{ for } 1\leq k\leq \ell-1\}.$ From $\mathcal{F}_{e_{i}}=\mathcal{R}_{e_{i}}-\mathcal{B}_{e_{i}}$ we can randomly determine the position of the coding element $e_i$ in the next cell. The positions in $\mathcal{F}_{e_{i}}$ maximize the distance between pixels, and guarantee choosing a different coding element than the first $\ell-1$ projections. For $e_2$, it would be the blue square highlighted in the next right cell.
	\\
	\textit{Step 4: Optimize across the $\ell$-th DOE.} Steps 2 and 3 are repeated for all coding elements $e_i$ of the admissible variable $d$. Thus, the spatial distribution of the $\ell$-th DOE is optimized.
	\\
	\textit{Step 5: Optimize across all DOEs.} The set of DOEs maximizes the variances along snapshots until $\ell=C_{d}$. If $\ell=C_{d}+1$, a new DOE dedicated to optimizing the spatial distribution must be generated. Thus, from $\ell=C_{d}+2$ to $\ell=2C_{d}$, the temporal and spatial correlation must be exploited considering the steps $1-4$, until the $L$-projections are completed.

	\subsubsection{End-to-end unfolding co-design of sensing matrix}
	The most important advantage of unfolding co-design is the ability to simultaneously design the entire or \textit{end-to-end} sensing process and the phase retrieval algorithm. Define the global noiseless measurement vector for the $k$-th diffraction zone $\boldsymbol{y}_{k}=[y_{k,1,1},\dots,y_{k,n,L}]^{T} \in \mathbb{R}^{m=nL}$ and consider the set of sensing matrices $\{\boldsymbol{A}_{k}\}_{k=1}^{3}$ with\par\noindent\small
	\begin{align}
		\boldsymbol{A}_{1} &= \left[ \overline{\boldsymbol{D}}_{1}\boldsymbol{F}\overline{\boldsymbol{T}}\boldsymbol{F}^{H},\cdots, \overline{\boldsymbol{D}}_{L}\boldsymbol{F}\overline{\boldsymbol{T}}\boldsymbol{F}^{H} \right]^{H} & \text{ (\textit{Near zone}) },\nonumber\\
		\boldsymbol{A}_{2} &= \left[ \overline{\boldsymbol{D}}_{1}\overline{\boldsymbol{Q}}\boldsymbol{F},\cdots, \overline{\boldsymbol{D}}_{L}\overline{\boldsymbol{Q}}\boldsymbol{F} \right]^{H} & \text{ (\textit{Middle zone}) },\nonumber \\
		\boldsymbol{A}_{3} &= \left[ \overline{\boldsymbol{D}}_{1}\boldsymbol{F}^{H},\cdots, \overline{\boldsymbol{D}}_{L}\boldsymbol{F}^{H} \right]^{H} & \text{ (\textit{Far zone}) },
		\label{eq:matrices}
	\end{align}\normalsize
	where $\boldsymbol{F} \in \mathbb{C}^{n\times n}$ is the discrete Fourier transform matrix and $\boldsymbol{T} \in\mathbb{C}^{n\times n}$ ($\boldsymbol{Q}\in\mathbb{C}^{n\times n}$) is the auxiliary orthogonal diagonal matrix that depends on the propagation distance (wavelength of the coherent source) to model the near (middle) zone (see \cite[Chapter~4]{poon2014introduction} for further details). The sensing process for the $k$-th diffraction zone is $\boldsymbol{y}_{k} = \lvert \boldsymbol{A}_{k}\boldsymbol{x} \rvert$, which corresponds to the first layer $f^{1}(\boldsymbol{x};\boldsymbol{A}_{k})=\lvert \boldsymbol{A}_{k}\boldsymbol{x} \rvert$ of the deep network (Figure \ref{fig:basicNetwork}) to model the CDP for the $k$-th diffraction zone.
	
	We seek to recover the signal of interest $\boldsymbol{x}$ from the embedded phaseless data $\boldsymbol{y}_{k}$ given the knowledge of $\boldsymbol{A}_{k}$ by solving the following optimization problem 
	\begin{align}
		\mathcal{P}_{0}: \hspace{1em} \text{ Find } \boldsymbol{x} \hspace{1em} \text{ subject to } \hspace{1em} \boldsymbol{y}_{k} = \lvert \boldsymbol{A}_{k}\boldsymbol{x} \rvert,
	\end{align}
	where the non-convex constraint models the measurement process. Additionally, if $\boldsymbol{x}^{\dagger}$ is a feasible point of the problem $\mathcal{P}_{0}$, then $e^{-j\varphi}\boldsymbol{x}^{\dagger}$ is also a solution to the problem for an arbitrary phase constant $\varphi$. Hence, it is only possible to recover the underlying signal of interest up to a constant global phase $\varphi \in \mathbb{R}$. This ambiguity arising from the global phase leads naturally to the following meaningful quantifying metric of closeness of the recovered signal $\boldsymbol{x}^{\dagger}$ to the true signal $\boldsymbol{x}$: 
	\begin{align}
		\textrm{dist}(\boldsymbol{x},\boldsymbol{x}^{\dagger}) = \min_{\varphi\in [0,2\pi)} \lVert \boldsymbol{x} - e^{-j\varphi}\boldsymbol{x}^{\dagger} \rVert_{2}.
		\label{eq:relError}
	\end{align}
	If $\textrm{dist}(\boldsymbol{x},\boldsymbol{x}^{\dagger})=0$, then $\boldsymbol{x}$ and $\boldsymbol{x}^{\dagger}$ are equal \emph{up to some global phase}.
	
	The end-to-end unfolding co-design seeks to jointly design the sensing matrix (encoder module or first layer in Figure \ref{fig:basicNetwork}) and the reconstruction algorithm (decoder module from layers 2 to $T$), in contrast to the existing methodologies that either consider the development of the reconstruction algorithm or the design of only DOEs. As a result, it is a unification of both approaches. Denote the class of decoder functions parametrized on a set of parameters $\Phi$ by $\mathcal{S}_{\Phi}:\mathbb{R}^{m}\rightarrow \mathbb{R}^{n}$. Assume that for a given measurement matrix $\boldsymbol{A}_{k}$ and the corresponding measurement vector $\boldsymbol{y}_{k}$, an estimation $\hat{\boldsymbol{x}}$ of the signal of interest is provided by a certain characterization of the decoder function, i.e.,
	\begin{align}
		\hat{\boldsymbol{x}} = \mathcal{S}_{\theta}(\boldsymbol{y}_{k};\boldsymbol{A}_{k}) \equiv \mathcal{S}_{\theta}(f^{1}(\boldsymbol{x};\boldsymbol{A}_{k});\boldsymbol{A}_{k})
	\end{align}
	where $\theta \in \Phi$ denotes a characterization of the decoder module from the original class. In addition, since $\boldsymbol{A}_{k}$ models a physical image formation model, it is desired to define the set of feasible sensing matrices for the optical setup as $\mathcal{T} \subset \mathbb{C}^{m\times n}$. This set contains the physical constraints related to the implementation of the DOE encapsulated in matrices $\boldsymbol{D}_{\ell}$ as in \eqref{eq:matrices}; see Section \ref{subsec:phy_cons} later for examples of these constraints. This, in consequence, means that a realistic $\boldsymbol{A}_{k}$ depends on the feasibility of DOE; see, for instance, Figure \ref{fig:systems} for a practical implementation. Then, for a fixed characterization of the decoder function $\mathcal{S}_{\theta \in \Phi}$, the sensing matrix $\boldsymbol{A}_{k}\in \mathcal{T}$ for the $k$-th diffraction zone is obtained by solving the optimization problem
	\begin{align}
		\minimize_{\boldsymbol{A}_{k}\in \mathcal{T}} ~\mathbb{E}\left[ \textrm{dist}(\boldsymbol{x};\mathcal{S}_{\theta}(\boldsymbol{y}_{k}\equiv f^{1}(\boldsymbol{x};\boldsymbol{A}_{k});\boldsymbol{A}_{k}))\right],
		\label{eq:designMatrix}
	\end{align}
	where the expectation is over the distribution of $\boldsymbol{x}$ through the chosen image dataset for end-to-end training. Note that this approach considers the signal reconstruction accuracy as a criterion with respect to a particular realization of the decoder functions. Thus, $\boldsymbol{A}_{k}$ obtained in this technique is task-specific encoding matrix which considers not only the underlying distribution of the signal but also the class of decoder functions resulting in a superior performance. After solving the aforementioned optimization problem, the sensing matrix is utilized for data-acquisition purposes while the fixed decoder module $\mathcal{S}_{\theta}$ is used for the reconstruction of $\boldsymbol{x}$. \vspace{-1em}
	
	\begin{figure}[t]
		\vspace{-1em}
		\centering
		\includegraphics[width=0.8\linewidth]{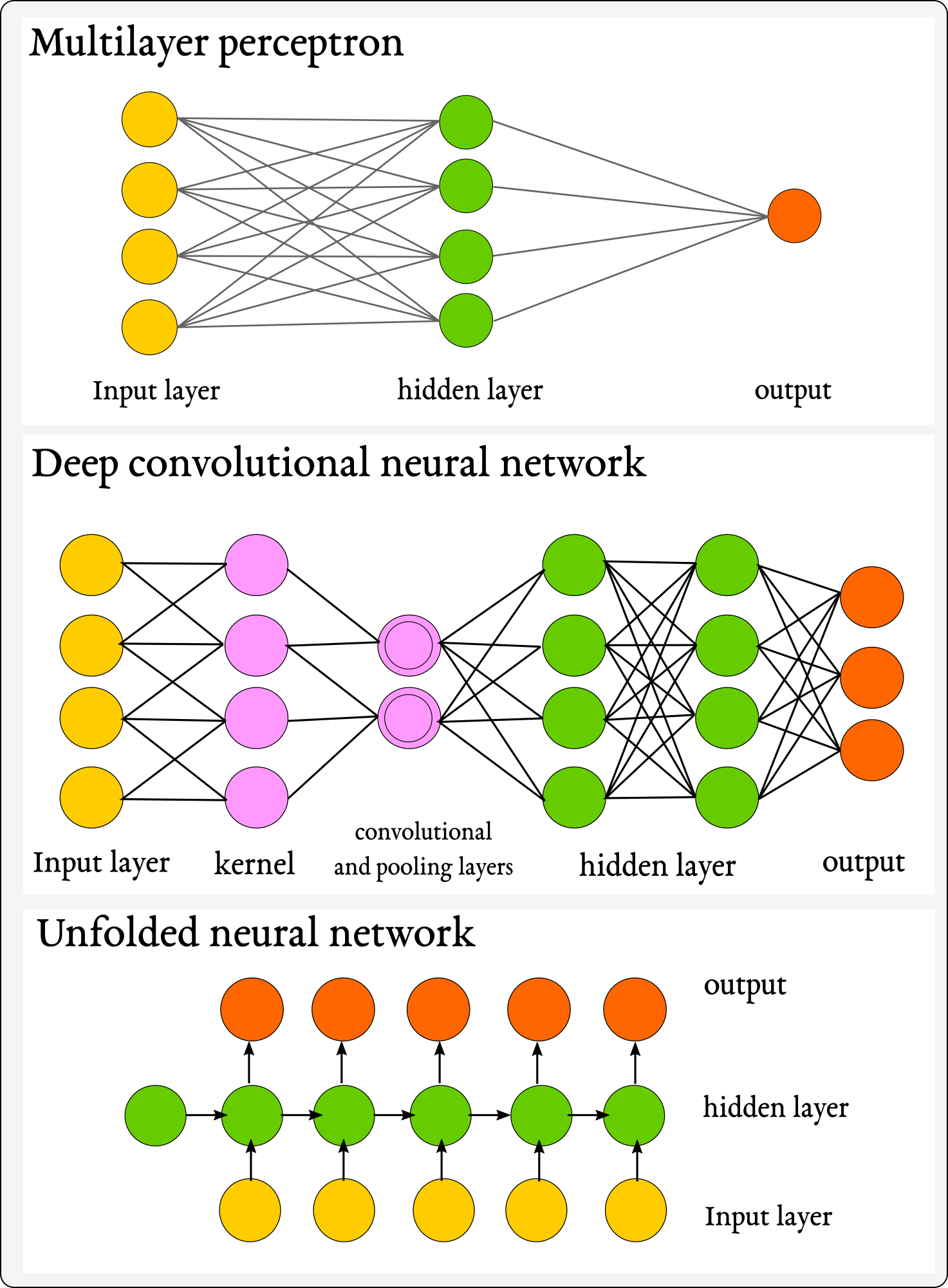}
		\caption{\small{Conventional network architectures that are popular in signal/image processing and computer vision applications. In a multilayer perceptron (top), all the neurons are fully connected. In a convolutional neural network (middle), the neurons are sparsely connected and the weights are shared among different neurons and convolutional layers. Unfolded neural network (bottom) is a recurrent structure where connections between nodes form a directed or undirected graph along a temporal sequence. }}\label{fig:ODI}
		\vspace{-1em}
	\end{figure}
	
	\subsection{DOE design with sparsity prior on the scene}
	In optical applications such as radio astronomy \cite{fienup1987phase} and microscopy \cite{kocsis2021ssr}, the scene $\boldsymbol{x}$ is naturally sparse in some orthogonal domain $\boldsymbol{\Psi}$, i.e. $\boldsymbol{\Psi}^{H}\boldsymbol{\Psi}=\boldsymbol{I}$; common examples include wavelet or discrete cosine transform). Then, there exists a \textit{s}-sparse, $s\ll n$, representation $\boldsymbol{z}\in\mathbb{C}^{n}$ such that $\|\boldsymbol{z}\|_{0}=s$, where $\boldsymbol{x}=\boldsymbol{\Psi}^{H}\boldsymbol{z}$ and $\|\cdot\|_{0}$ is the $\ell_{0}$ pseudo-norm. It was shown in \cite{guerrero} that very low estimation errors in sparse representation $\boldsymbol{z}$ of $\boldsymbol{x}$ were obtained when the admissible random variable $d$ that models the DOEs satisfies $\mathbb{E}[d]\neq0$. This result suggests the type of coding elements of $d$ needed for highest performance without any restriction on the design strategy. Consequently, the unfolding co-design method is also valid for the sparsity prior scenarios; only the first layer needs to be changed to incorporate the aforementioned result by choosing an appropriate $d$. Note that $d$ plays a crucial role to uniquely identify $\boldsymbol{x}$ from CDP and the flexibility of the unfolding framework to combine model-based theoretical results and neural networks. Thus, it is desirable to consider and replicate \cite{guerrero} for analogous analysis of a given application to achieve optimal imaging quality.
	
	\subsection{Physical constraints}
	\label{subsec:phy_cons}
	The optimization problem in \eqref{eq:designMatrix} is intended to design the measurement matrix $\boldsymbol{A}_{k}$ for a given diffraction zone $k$. More physical constraints may be incorporated in the problem. For example, given the block-matrix structure in \eqref{eq:matrices}, the DOE may be modelled as a piecewise continuous function \cite{pinilla2022hybrid}, wherein the constant intervals of the function are characterized by the random variable $d$. Similarly, the Fresnel order or DOE thickness may be described using a smoothing function \cite{rostami2021power}. In each one of these cases, the physical parameters are learnable.
	
	\textit{Fresnel order (thickness):} The thickness of DOE is defined as $Q=2\pi m_{Q}$ (in radians), where $m_{Q}$ is the ``Fresnel order" of the mask that is not necessarily an integer. Denote the wavelength of the source by $\lambda_{0}$ and the resulting phase profile of DOE by $\boldsymbol{\varphi}_{\lambda_{0}}$. Then, the phase profile of the DOE considering the thickness is given by
	\begin{equation}
		(\hat{\boldsymbol{\varphi}}_{\lambda_{0}})_{p,q} = \textrm{mod}((\boldsymbol{\varphi}_{\lambda_{0}})_{p,q} + Q/2,Q)-Q/2,
		\label{lens4}
	\end{equation}
	resulting in $\hat{\boldsymbol{\varphi}}_{\lambda _{0}}$ taking values in the interval $[-Q/2$, $Q/2)$.
	
	\textit{Number of Levels: } The DOE is defined on a discretized 2-D grid. We obtain a piece-wise invariant surface for DOE after non-linear transformation of its absolute phase. The uniform grid discretization of the wrapped phase profile $\hat{\boldsymbol{\varphi}}_{\lambda _{0}}$ up to $N$ levels is
	\begin{equation}
		(\boldsymbol{\beta}_{\lambda_{0}})_{p,q} =\lfloor (\hat{\boldsymbol{\varphi}}_{\lambda_{0}})_{p,q} /N \rfloor \cdot N\text{,}
		\label{lens5}
	\end{equation}
	where $\lfloor w \rfloor$ denotes floor operation that yields the largest integer less than or equal to $w$. The values of $\boldsymbol{\beta}_{\lambda_{0}}$ are restricted to $[-Q/2$, $Q/2)$, where each step of the piecewise function is a coding element of $d$. Here, $Q$ is an upper bound of thickness phase of $\boldsymbol{\beta}_{\lambda_{0}}$.
	
	The floor and modulo functions are not differentiable. Therefore, a smoothing approximation is used for optimizing the DOE thickness and number of levels; see \cite{rostami2021power} for details.
	
	\textit{Pixel resolution and size of DOE:} These parameters are also considered while designing the measurement matrix because, in practice, all sizes and resolutions may not be implementable \cite{dun2020learned}. Further details on the impact of these physical DOE constraints in a practical imaging application are available in~\cite{rostami2022design}. \vspace{-0.5em}
	
	\section{Unfolding for Optical Imaging}
	Compared to some popular neural network architectures (Figure~\ref{fig:ODI}), the unfolded network is a directed acyclic graph that is trained like any standard neural network. However, the input to this unfolded network is not a single vector from the sequence but rather it is the entire sequence used all at once. The output used to compute the gradients is also the entire sequence of output values that a typical neural network would have produced for each step in the input sequence. As such the unfolded network has multiple inputs and multiple outputs with identical weight matrices for each step.
	
	A preliminary step in unfolding is to mathematically formulate a proper parametrized class of decoder and encoder modules that facilitate incorporating the domain knowledge. To this end, the decoder is obtained by first formulating phase retrieval problem as a minimization program and then resorting to first-order optimization techniques that lay the ground for creating a rich class of parametrized decoder functions. Once such a class is formalized, a connection between deep neural networks and first-order optimization techniques is established. \vspace{-1em}
	
	\subsection{Underlying principles}
	\label{subsec:princi}
	Assume $\mathcal{S}_{\theta_{1}\in \Phi}$ and $\mathcal{S}_{\theta_{2}\in \Phi}$ represent two realizations of the decoder module from the same class. Denote the fixed sensing matrix that will be the solution to \eqref{eq:designMatrix} for $\theta = \theta_{2}$ by $\boldsymbol{A}_{k}$ for the $k$-th diffraction zone. Then, for a signal of interest $\boldsymbol{x}$ and corresponding phaseless measurements $\boldsymbol{y}_{k}$, it is possible that
	\begin{align}
		\mathbb{E}\left[ \textrm{dist}(\boldsymbol{x};\mathcal{S}_{\theta_{1}}(\boldsymbol{y}_{k};\boldsymbol{A}_{k}))\right] \leq \mathbb{E}\left[ \textrm{dist}(\boldsymbol{x};\mathcal{S}_{\theta_{2}}(\boldsymbol{y}_{k};\boldsymbol{A}_{k}))\right].
	\end{align}
	As a result, it is more appropriate to consider the design of the sensing matrix $\boldsymbol{A}_{k}$ with respect to $\mathcal{S}_{\theta_{1}}$ and find an optimal $\theta^{*}\in \Phi$ such that\par\noindent\small
	\begin{align}
		\mathbb{E}\left[ \textrm{dist}(\boldsymbol{x};\mathcal{S}_{\theta^{*}}(\boldsymbol{y}_{k};\boldsymbol{A}_{k}))\right] \leq \mathbb{E}\left[ \textrm{dist}(\boldsymbol{x};\mathcal{S}_{\theta}(\boldsymbol{y}_{k};\boldsymbol{A}_{k}))\right], \forall ~ \theta \in \Phi.
	\end{align}\normalsize
	
	If such a characterization $\theta^{*}$ is known, the sensing matrix is easily designed. The above observation is further evidence for the paramount importance of developing a unified framework that allows for a joint optimization of the reconstruction algorithm (i.e. finding an optimal or sub-optimal $\theta^{*}$) and the sensing matrix. Then, the problem of jointly designing $\boldsymbol{A}_{k}$ and the reconstruction algorithm is equivalent to the optimization program
	\begin{align}
		\minimize_{\boldsymbol{A}_{k} \in \mathcal{T}, \theta\in \Phi} \mathbb{E}\left[ \textrm{dist}(\boldsymbol{x};\mathcal{S}_{\theta}(f^{1}(\boldsymbol{x};\boldsymbol{A}_{k});\boldsymbol{A}_{k}))\right].
	\end{align}
	Next, we describe tackling this optimization problem using the available data.
	\\
	\textbf{Encoder architecture:} Define the hypothesis class $\mathcal{H}_{e}$ of the possible encoder functions (module) whose computation dynamics mimics the behaviour of the data-acquisition system in a phase retrieval model (e.g. $\mathbb{E}[d]\neq0$, temporal or spatial correlation of $\boldsymbol{A}_{k}$). Then, 
	\begin{align}
		\mathcal{H}_{e} = \{f^{1}(\cdot;\boldsymbol{A}_{k}): \boldsymbol{x}\rightarrow \lvert \boldsymbol{A}_{k}\boldsymbol{x} \rvert: \boldsymbol{A}_{k}\in \mathbb{C}^{m\times n} \}.
	\end{align}
	The superscript in $f^{1}$ denotes that this hypothesis class corresponds to the first layer of the unfolded network. It can be interpreted as a \textit{one-layer neural network} with $n$ input neurons and $m$ output neurons, where the activation function is $\lvert \cdot \rvert$ with the trainable weight matrix $\boldsymbol{A}_{k}$. The encoder (matrix $\boldsymbol{A}_{k}$) for a given realization of a decoder function $\mathcal{S}_{\theta}$ with respect to the hypothesis class $\mathcal{H}_{e}$ is designed by solving the following \textit{learning problem} \par\noindent\small
	\begin{align}
		\minimize_{f^{1}\in \mathcal{H}_{e}} \mathbb{E}\left[ \textrm{dist}(\boldsymbol{x};\mathcal{S}_{\theta}\circ f^{1}(\boldsymbol{x};\boldsymbol{A}_{k}))\right],
	\end{align}\normalsize
	where $\circ$ stands for composition of functions. This problem seeks to find an encoder function $f^{1}(\cdot;\boldsymbol{A}_{k})$ such that the resulting functions maximizes the reconstruction accuracy according to the decoder function $\mathcal{S}_{\theta}$.
	\\
	\textbf{Decoder architecture:} The goal here is to obtain an estimate of the underlying signal of interest from the phaseless measurements of the form $f^{1}(\boldsymbol{x};\boldsymbol{A}_{k})=\lvert \boldsymbol{A}_{k}\boldsymbol{x} \rvert$, for a given sensing matrix $\boldsymbol{A}_{k}$ of the $k$-th diffraction zone. Equivalently, a decoder function seeks to tackle the following problem\par\noindent\small
	\begin{align}
		\minimize_{\boldsymbol{z}\in \mathbb{C}^{n}} \hspace{1em}\frac{1}{2m} \left\lVert f^{1}(\boldsymbol{z};\boldsymbol{A}_{k}) - f^{1}(\boldsymbol{x};\boldsymbol{A}_{k}) \right\rVert_{2}^{2} \hspace{1em} \text{ subject to } \hspace{1em} \boldsymbol{z}\in \Gamma,
	\end{align}\normalsize
	where $\Gamma$ represents the search space of the underlying signal of interest. For instance, if $\boldsymbol{x}$ is $s$-sparse, this information is encoded in $\Gamma=\{\boldsymbol{z}\in \mathbb{C}^{n}: \lVert \boldsymbol{z}\rVert_{0}=s\}$. \vspace{-0.5em}
	
	\begin{strip}
		\vspace{-1em}
		\begin{tcolorbox}[colback=infocolor,title={\textbf{UPR without priors}},boxsep=1pt,left=4pt,right=4pt,top=2pt,bottom=2pt]
			\begin{multicols}{2}
				Without sparsity prior, the optimization problem for phase retrieval follows directly as \par\noindent\small
				\begin{align}
					\minimize_{\boldsymbol{z}\in \mathbb{C}^{n}} \mathcal{L}\left(\boldsymbol{z};f^{1}(\boldsymbol{x};\boldsymbol{A}_{k})\right) \equiv \frac{1}{2m} \left\lVert f^{1}(\boldsymbol{x};\boldsymbol{A}_{k}) -f^{1}(\boldsymbol{z};\boldsymbol{A}_{k}) \right\rVert_{2}^{2},
					\label{eq:optiProblem}
				\end{align}\normalsize
				where $m=nL$, $L$ is the number of snapshots of the scene, and signal length $n$ is a constant parameter that controls the step-size of each iteration.
				
				The iterations for finding the critical points of this non-convex problem are derived as follows. Starting from a proper initial point $\boldsymbol{z}^{(0)}$, a sequence of points $\{\boldsymbol{z}^{(0)},\boldsymbol{z}^{(1)}, \boldsymbol{z}^{(2)},\dots,\boldsymbol{z}^{(T)} \}$ is generated according to the following update rule:
				\begin{align}
					\boldsymbol{z}^{(t+1)} = \boldsymbol{z}^{(t)} - \frac{1}{m}\boldsymbol{G}^{(t)}\nabla_{\boldsymbol{z}} \mathcal{L}(\boldsymbol{z}^{(t)};f^{1}(\boldsymbol{x};\boldsymbol{A}_{k})).
				\end{align}
				This update rule was shown to be highly effective in solving the phase retrieval problem \cite{zhang2016reshaped,wang2017sparse}. Therefore, it is instructive to further explore its potential in the unfolding context. The gradient of the objective function is
				\begin{align}
					&\nabla_{\boldsymbol{z}} \mathcal{L}(\boldsymbol{z}^{(t)};f^{1}(\boldsymbol{x};\boldsymbol{A}_{k})) \nonumber\\
					&= \boldsymbol{A}_{k}^{H}\left(\boldsymbol{A}_{k}\boldsymbol{z}^{(t)} - f^{1}(\boldsymbol{x};\boldsymbol{A}_{k})\odot \textrm{Ph}(\boldsymbol{A}_{k}\boldsymbol{z}^{(t)})\right),
				\end{align}
				where $\odot$ is the inner product and the function $\textrm{Ph}(\boldsymbol{z})$ is applied element-wise and captures the phase of the vector argument; e.g., for real valued signals $\textrm{Ph}(\boldsymbol{z}) = \textrm{sign}(\boldsymbol{z})$.
			\end{multicols}
			
			\begin{center}
				\includegraphics[width=0.95\linewidth]{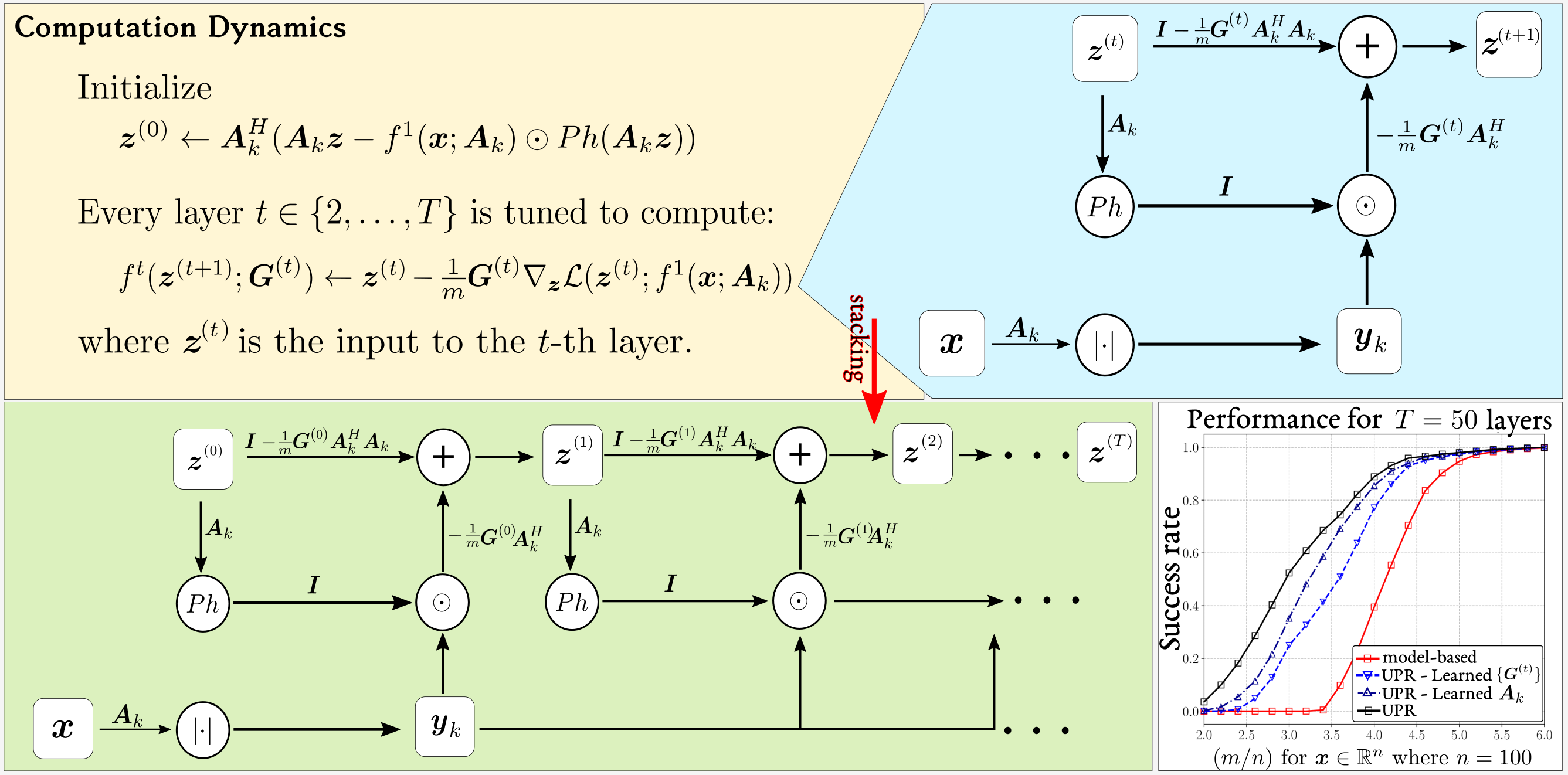}
				\vspace{-0.5em}
				\captionof{figure}{Single iteration (top left) of the UPR method executes both linear and nonlinear operations and is thus recast into a network layer (top right). Stacking the layers together, a deep unfolded network (bottom left) is formed. The network is subsequently trained using paired inputs and outputs by backpropagation to optimize the parameters $\boldsymbol{G}^{(t)}$ and $\boldsymbol{A}_{k}$. The learning rate was set to $10^{-4}$ for $100$ epochs, using Adam stochastic optimizer. The empirical success rate (bottom right), obtained by averaging over $100$ Monte-Carlo trials, is shown with respect to the measurement-to-signal-length ratio $(m/n)$, when the signal length is fixed and set to $n = 100$. The size of the training dataset is $2048$, where each data point is a $n$-dimensional vector following a standard Gaussian distribution. In the model-based case, the sensing matrix $\boldsymbol{A}_{k}$ follows a standard Gaussian distribution with variance equal to one.}
				\label{fig:unrolling1}
			\end{center} 
			
		\end{tcolorbox}
		\vspace{-1em}
	\end{strip}
	
	\subsection{Application to phase retrieval}
	Deep unfolded networks have repeatedly displayed great promise in various signal processing applications \cite{morales2021object,morales2022learning} by exploiting copious amount of data along with the domain knowledge gleaned from the underlying problem. This is suitable to deploy for phase retrieval in optical imaging, particularly in non-convex settings where bounding the complexity of signal processing algorithms is often traded-off for performance. In general, iterative optimization techniques are a popular choice for non-convex phase retrieval problems. Specifically, first-order methods are widely used iterative optimization techniques because of their low per-iteration complexity and efficiency in complex scenarios. One such prominent approach that is also suitable for training the decoder is the projected gradient descent (PGD) algorithm. Here, given an initial point $\boldsymbol{z}^{(0)}$, the $t$-th iteration of PGD is
	\begin{align}
		\boldsymbol{z}^{(t+1)} = \mathcal{P}_{\Gamma}\left(\boldsymbol{z}^{(t)} - \frac{\alpha^{(t)}}{m}\nabla_{\boldsymbol{z}}\ell(\boldsymbol{z}^{(t)})\right),
	\end{align}
	where $\mathcal{P}_{\Gamma}: \mathbb{C}^{n}\rightarrow \Gamma \subseteq \mathbb{C}^{n}$ denotes a mapping function of the vector argument to the feasible set $\Gamma$, $\nabla_{\boldsymbol{z}}\ell(\boldsymbol{z}^{(t)})$ is the gradient of the objective function obtained at the point $\boldsymbol{z}^{(t)}$, and $\alpha^{(t)}$ represents the step-size of the PGD algorithm at the $t$-th iteration. To bound the computational cost of first-order methods, a sensible approach is to fix the total number of iterations of such algorithms (e.g., $T$), and to follow up with a proper choice of the parameters for each iteration that results in the best improvement in the objective function, while allowing only $T$ iterations.
	
	Define a parametrized mapping operator $f^{t}(\cdot;\boldsymbol{\rho}^{t}): \mathbb{C}^{n}\rightarrow \mathbb{C}^{n}$ for $t\geq 2$ (see Figure \ref{fig:basicNetwork}) as 
	\begin{align}
		f^{t}(\boldsymbol{z};\boldsymbol{\rho}^{t}) = \mathcal{P}_{\Gamma}\left(\boldsymbol{z} - \frac{1}{m}\boldsymbol{G}^{(t)}\nabla_{\boldsymbol{z}}\ell(\boldsymbol{z})\right),
		\label{eq:preconditioning}
	\end{align}
	where $\boldsymbol{\rho}^{t}= \{\boldsymbol{G}^{(t)}\}$ is the set of parameters of the mapping function $f^{t}(\cdot;\boldsymbol{\rho}^{t})$, and $\boldsymbol{G}^{(t)}$ is a positive semi-definite matrix. This mapping is then utilized to model various first-order optimization techniques for a given problem. In particular, for the PGD algorithm, performing $T-1$ iterations of the above form (Figure \ref{fig:basicNetwork}) is modelled as\par\noindent\small
	\begin{align}
		\mathcal{N}_{\Gamma}^{T-1}(\boldsymbol{x}) = f^{T}(\cdot;\boldsymbol{\rho}^{T}) \circ f^{T-1}(\cdot;\boldsymbol{\rho}^{T-1}) \circ \cdots \circ f^{2}(\cdot;\boldsymbol{\rho}^{2}),
	\end{align}\normalsize
	where the above function corresponds to the PGD in all scenarios, including for a fixed step-size $\boldsymbol{G}^{(t)}=\alpha\boldsymbol{I}$, time-varying step-sizes $\boldsymbol{G}^{(t)}=\alpha^{(t)}\boldsymbol{I}$, or the general preconditioned PGD algorithm with an arbitrary choice of $\boldsymbol{G}^{(t)}\succeq 0$. 
	
	The specific preconditioning-based update rule in \eqref{eq:preconditioning} is known to be advantageous in model-based algorithms to speed-up the convergence rate \cite[Chapter 5]{nocedal1999numerical} and yield robustness when the sensing matrix is ill-conditioned (i.e., condition number is far greater than 1) \cite{jia2015preconditioning,wauthier2013comparative,kelner2022power}. This latter advantage is critical in practical imaging applications (listed in Table~\ref{tab:1}) to overcome imperfect system modelling, reconstruction artefacts such as mismatch, calibration errors, and even poorer performance for single snapshot of experimental sensing matrix~\cite{mojica2021high,mejia2018binary}. In the context of unfolding \cite{chen2018theoretical}, the benefits of preconditioning appear indirectly because it was learnt as a matrix that encapsulates both sensing (via its transpose) and preconditioning matrices.
	
	Finally, for a fixed $\boldsymbol{A}_{k}$, learning the deep decoder module with respect to a realization of the encoder function corresponds to the problem 
	\begin{align}
		\minimize_{\Gamma = \{\boldsymbol{G}^{(t)}\succeq 0\}} \mathbb{E}[\textrm{dist}(\boldsymbol{x};\mathcal{N}^{T-1}_{\Gamma} \circ f^{1}(\boldsymbol{x};\boldsymbol{A}_{k}))].
	\end{align}
	This formulation is akin to training a $T$-layer deep neural network $\mathcal{N}^{T-1}_{\Gamma} \circ f^{1}(\boldsymbol{x};\boldsymbol{A}_{k})$ with a task-specific architecture whose computation dynamics at each layer mimic the behaviour of one iteration of a first-order optimization algorithm, and the input to such a model-based deep scheme is given by a realization of the encoder module. Since domain knowledge is included in the design of such a deep unfolded network, it is more interpretable and has fewer training parameters than its black-box counterpart.\vspace{-0.5em}
	
	\begin{figure}[ht]
		\centering
		\includegraphics[width=0.9\linewidth]{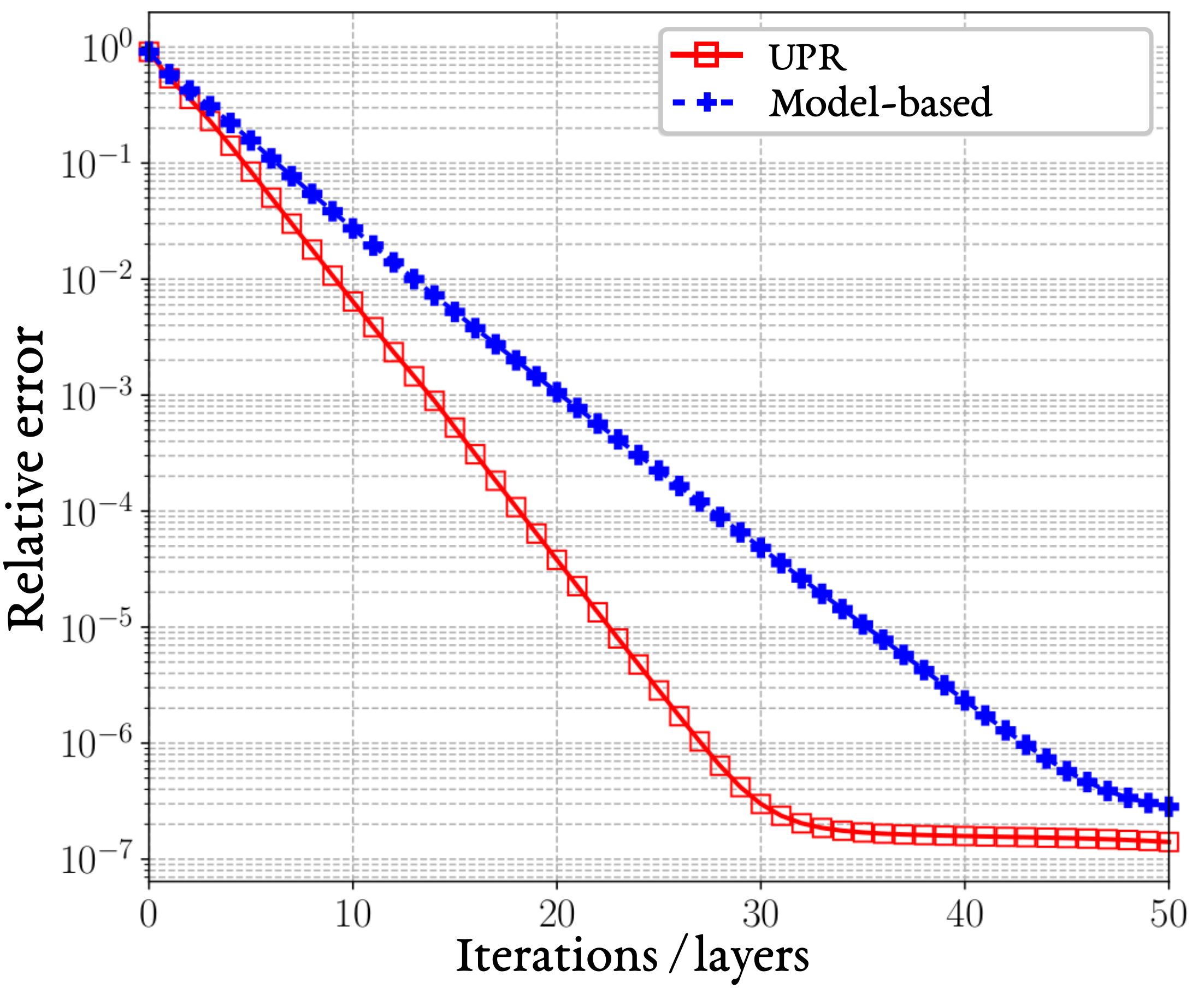}
		\captionof{figure}{Convergence behaviour of UPR \cite{naimipour2020unfolded} and model-based \cite{zhang2016reshaped} algorithms for $\boldsymbol{x}\in \mathbb{R}^{n}$, $n=100$, and $m=6n$. This experiment follows the same setup of Figure \ref{fig:unrolling1} in terms of step-size, learning rate, dataset and sensing matrices.}
		\vspace{-1em}
		\label{fig:iterUPRWithout}
	\end{figure}
	
	\section{Unfolding-Aided Techniques for Coded DOI}
	Prevailing usage of unfolding algorithms in phase retrieval is centered on the assumption or lack thereof sparsity in the target scenario in non-convex setting. More recently, optical applications such as 3-D object detection and 3-D phase imaging have also exploited unfolded networks. \vspace{-0.5em}
	
	\subsection{UPR without priors}
	The computational dynamics and network structure of UPR architecture (see box) for this scenario are depicted in Figure~\ref{fig:unrolling1}. From the empirical success rate it follows that the unfolded network significantly outperforms the model-based algorithm in all scenarios. 
	
	When the measurement matrix is fixed (by the physics of the system) and known \textit{a priori}, learning the preconditioning matrices $\boldsymbol{G}^{(t)}$ significantly improves the performance of the underlying signal recovery algorithm. This supports the proposition that a judicious design of the preconditioning matrices, when the number of layers (iterations) are fixed, indeed results in an accelerated convergence. The unfolding approach is better even when only the measurement matrix is learned while employing a fixed scalar step-size. With both learned $\boldsymbol{A}_{k}$ and $\boldsymbol{G}^{(t)}$, the unfolded network is the most successful.
	
	Figure~\ref{fig:iterUPRWithout} demonstrates the relative error (as in \eqref{eq:relError}) as a function of iterations (layers). It follows that UPR methodology benefits from accelerated convergence and an optimal point of the objective function is reached with as few as $T=30$ layers. This reveals that the proposed architecture may be truncated to even fewer iterations thereby saving the computations in the overall algorithm.
	
	\subsection{UPR with sparsity prior}
	\label{subsec:withprior}
	In sparse phase retrieval (see box), decoding signal $\boldsymbol{x}$ from the measurement vector $f^{1}(\boldsymbol{x};\boldsymbol{A}_{k})$ is formulated as a non-convex optimization problem (see Section~\ref{subsec:princi}). Both the objective function $\mathcal{L}(\boldsymbol{z};f^{1}(\boldsymbol{x};\boldsymbol{A}_{k}))$ and the constraint for this sparse problem are non-convex making it NP-hard in its general form. Figure~\ref{fig:unrolling2} shows the relevant UPR architecture, which significantly outperforms the standard state-of-the-art sparse model-based algorithm under various learned parameters. When the measurement matrix is imposed by the physical system, then learning the preconditioning matrices significantly increases the recovery performance. Considering learning only the matrix $\boldsymbol{A}_{k}$ and employing fixed step-sizes further indicates the effectiveness of learning task-specific measurement matrices tailored to the application at hand. 
	
\begin{figure}[t]
		\centering
		\includegraphics[width=0.9\linewidth]{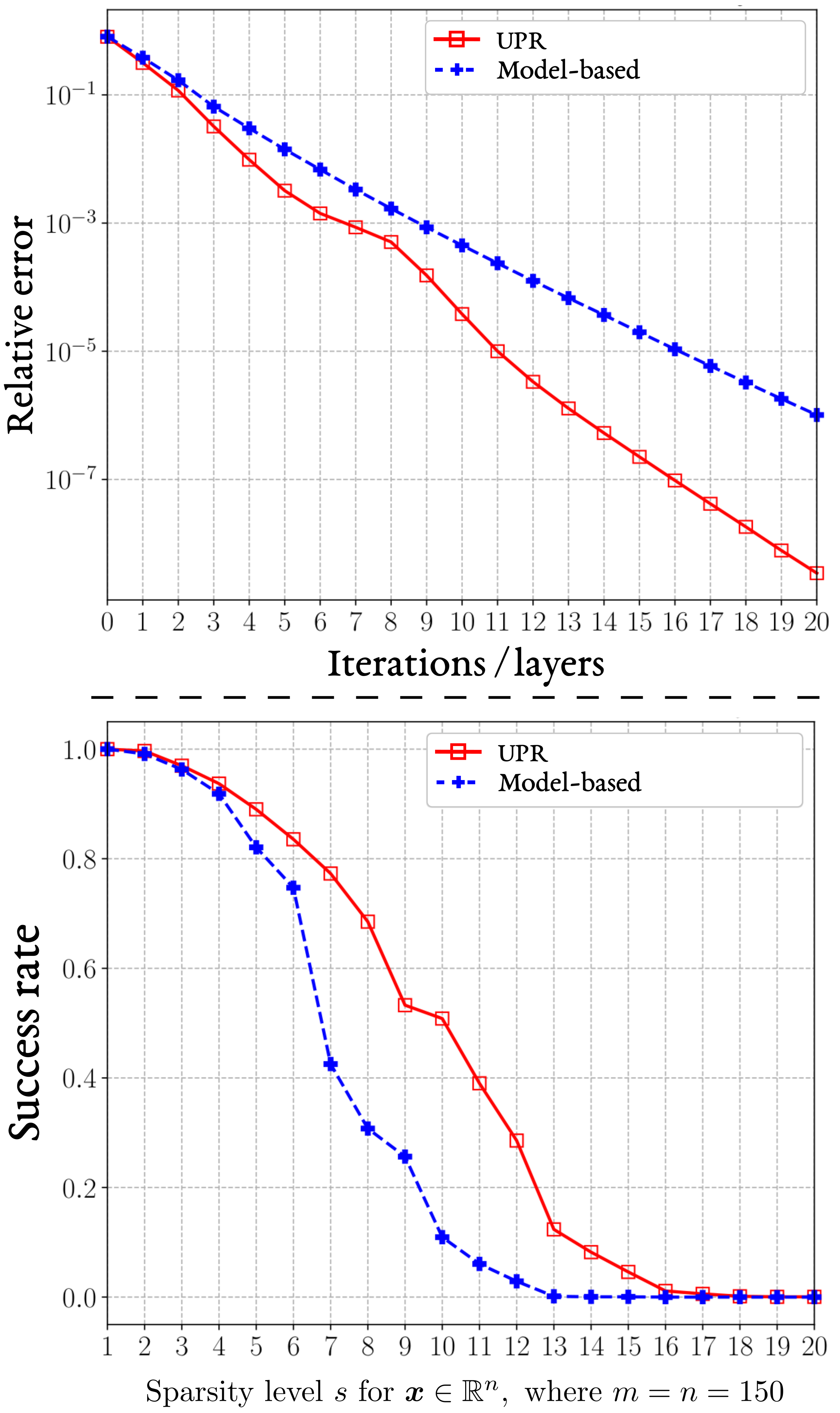}
		\captionof{figure}{Top: Convergence behaviour of UPR \cite{zhang2016reshaped} and model-based \cite{wang2017sparse} algorithms for $\boldsymbol{x}\in \mathbb{R}^{n}$, $n=m=300$. The step-size, learning rate, dataset and sensing matrices are as in Figure~\ref{fig:unrolling2}. Bottom: Empirical success rate versus the sparsity level $s=5$ for $\boldsymbol{x}\in \mathbb{R}^{n}$, $n=m=150$. }
		\vspace{-1em}
		\label{fig:sparseIterations}
	\end{figure}

	The objective of learning the preconditioning matrices is to accelerate the convergence. This may be validated through a per-layer analysis of sparse UPR framework in terms of achieved relative error (as in \eqref{eq:relError}). The model-based nature of the proposed approach facilitates analysis of the resulting interpretable deep architectures as opposed to the conventional black-box data-driven approaches. Figure \ref{fig:sparseIterations} (top) demonstrates the relative error versus number of iterations/layers for the case of recovering a $5$-sparse signal $\boldsymbol{x}\in \mathbb{R}^{n}$, $n = m = 300$. It follows that the UPR yields both lower relative error and faster convergence than the standard model-based algorithm \cite{zhang2016reshaped}. The success rate versus sparsity plot in Figure \ref{fig:sparseIterations} (bottom) further highlights the superior performance of UPR.
	
	\begin{strip}
		\vspace{-2em}
		\begin{tcolorbox}[colback=infocolor,title={\textbf{UPR with sparsity prior}},boxsep=1pt,left=4pt,right=4pt,top=2pt,bottom=2pt]
			
			\begin{multicols}{2}
				Starting from an initial guess $\boldsymbol{z}^{(0)}$, the algorithm estimates the signal in an iterative manner via update equations as
				\begin{align}
					\boldsymbol{z}^{(t+1)} = \mathcal{P}_{s}\left(\boldsymbol{z}^{(t)}-\frac{\alpha}{m}\nabla_{\boldsymbol{z}} \mathcal{L}_{tr}(\boldsymbol{z}^{(t)};f^{1}(\boldsymbol{x};\boldsymbol{A}_{k}))\right).
				\end{align}
				These iterations are viewed as performing PGD with a per-iteration step-size $\alpha$ on the truncated loss function objective $\mathcal{L}_{tr}$. The gradient is 
				\begin{align}
					&\nabla_{\boldsymbol{z}} \mathcal{L}_{tr}(\boldsymbol{z}^{(t)};f^{1}(\boldsymbol{x};\boldsymbol{A}_{k})) \nonumber\\
					&= \sum_{i\in \mathcal{I}_{t}}\left(\boldsymbol{a}_{k,i}^{H}\boldsymbol{z}^{(t)} - [f^{1}(\boldsymbol{x};\boldsymbol{A}_{k})]_{i}\frac{\boldsymbol{a}_{k,i}^{H}\boldsymbol{z}^{(t)}}{\lvert \boldsymbol{a}_{k,i}^{H}\boldsymbol{z}^{(t)} \rvert}\right)\boldsymbol{a}_{k,i},
				\end{align}
				where $\mathcal{I}_{t} = \left\{1\leq i\leq m: \lvert \boldsymbol{a}_{k,i}^{H}\boldsymbol{z}^{(t)} \rvert\geq \frac{[f^{1}(\boldsymbol{x};\boldsymbol{A}_{k})]_{i}}{1+\tau}\right\}$, the constant $\tau$ represents the truncation threshold, $\mathcal{P}_{s}(\boldsymbol{u}): \mathbb{C}^{n}\rightarrow \mathbb{C}^{n}$ sets all entries of $\boldsymbol{u}$ to zero except for the $s$ entries with the largest magnitudes. This truncation addresses the non-continuity of the gradient by eliminating the terms that could lead to unbounded descent direction \cite{cande2}. This provides a strong motivation to investigate this procedure for unfolding.
				
				The vector $\boldsymbol{z}^{(0)}$ is initialized as $\sqrt{\sum_{i=1}^{m} \frac{[f^{1}(\boldsymbol{x};\boldsymbol{A}_{k})]^{2}_{i}}{m}}\hat{\boldsymbol{z}}^{(0)}$ where $\hat{\boldsymbol{z}}^{(0)}\in \mathbb{C}^{n}$ is created by placing zeros in $\hat{\boldsymbol{z}}^{(0)}_{\hat{\mathcal{S}}}$ where the indices are not in $\hat{\mathcal{S}}$ (indices with the $s$ largest values of $\hat{\boldsymbol{z}}^{(0)}$). The principal eigenvector $\hat{\boldsymbol{z}}^{(0)}_{\hat{\mathcal{S}}}\in \mathbb{C}^{s}$ is determined by performing power method iterations on the matrix
				\begin{align}
					\boldsymbol{\Lambda} = \frac{1}{\lvert \mathcal{I}_{0} \rvert} \sum_{i\in \mathcal{I}_{0}} \frac{\boldsymbol{a}_{k,i,\hat{\mathcal{S}}}\boldsymbol{a}_{k,i,\hat{\mathcal{S}}}^{H}}{\lVert \boldsymbol{a}_{k,i,\hat{\mathcal{S}}} \rVert_{2}^{2}}.
				\end{align}
				The condition $\mathbb{E}\left[\sum_{\ell=1}^{L}\overline{\boldsymbol{D}}_{\ell}\boldsymbol{D}_{\ell}\right]=r\boldsymbol{I}$ imposed over the set of DOEs to guarantee uniqueness is needed for this initialization to return an accurate estimation of $\boldsymbol{x}$ \cite{guerrero}. This result connects the initialization procedure with the recovery conditions of CDP phase retrieval. 
			\end{multicols}
			
			\begin{center}
				\includegraphics[width=0.95\linewidth]{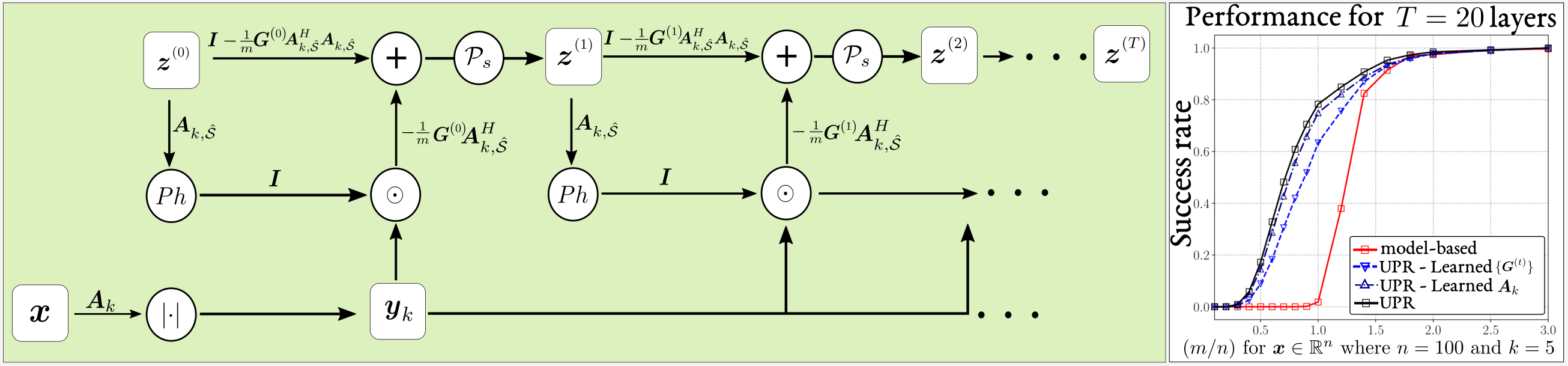}
				\vspace{-0.5em}
				\captionof{figure}{The unfolded network (left) is trained using paired inputs and outputs by backpropagation to optimize the parameters $\boldsymbol{G}^{(t)}$ and $\boldsymbol{A}_{k}$. Here, the signal length $n$ is a constant that controls the step-size (= 1) of each iteration. The learning rate was set to $10^{-4}$ for $100$ epochs, using Adam stochastic optimizer. The empirical success rate (right), obtained by averaging over $100$ Monte-Carlo trials, versus the measurement-to-signal-length ratio $(m/n)$, when the signal length is set to $n = 100$ and $s=5$. The size of the training dataset is $2048$, where each data point is a $n$-dimensional vector following a standard Gaussian distribution, whose $(n-s)$ entries set to zero are chosen uniformly at random. In the model-based case, the sensing matrix $\boldsymbol{A}_{k}$ follows a standard Gaussian distribution with variance equal to one.}
				\label{fig:unrolling2}
			\end{center} 
		\end{tcolorbox}
		\vspace{-0.5em}
	\end{strip}
	
	\subsection{Unfolded object detection}
\begin{figure}[t!]
		\centering
		\includegraphics[width=0.9\linewidth]{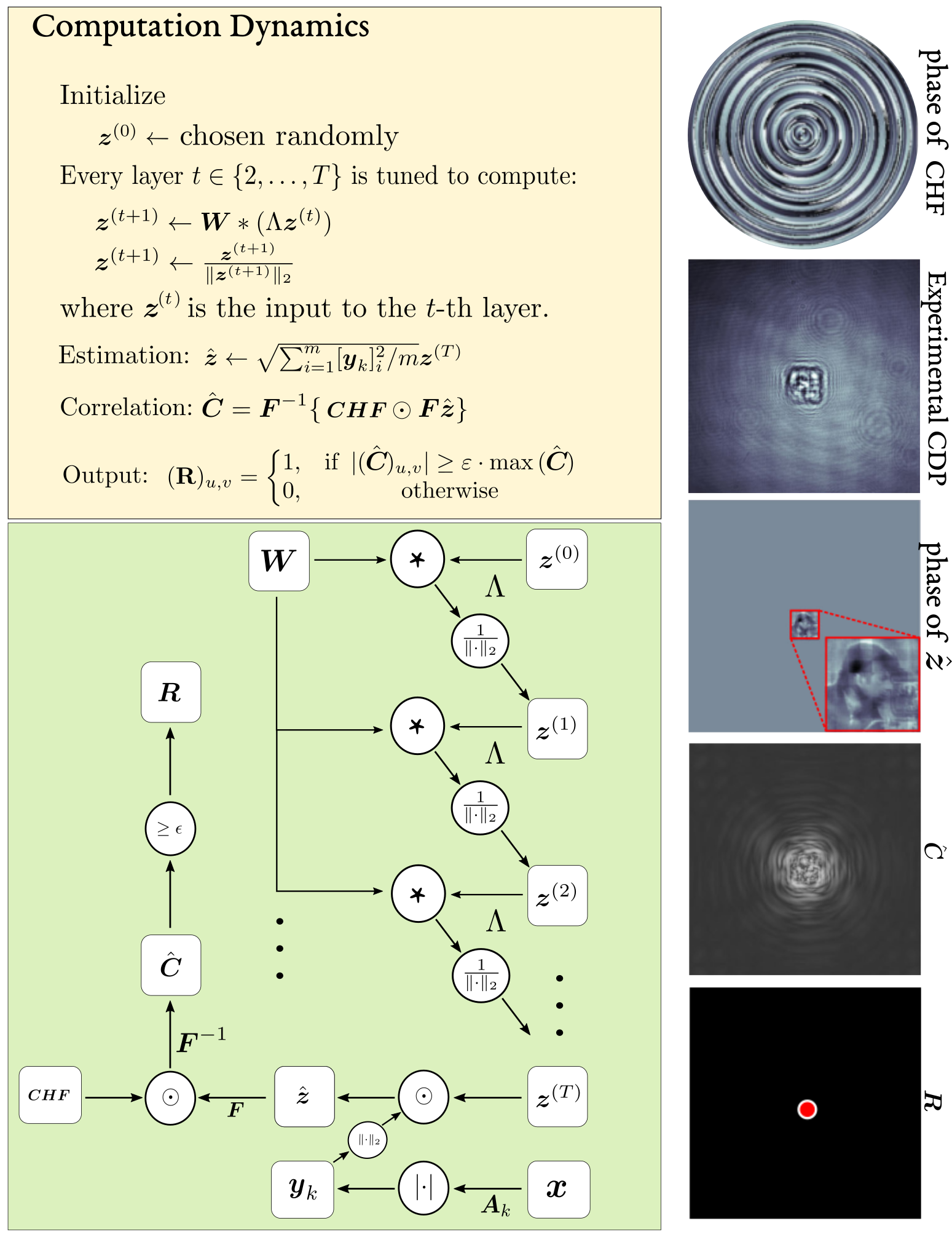}
		\vspace{-0.5em}
		\captionof{figure}{Top left: Algorithm for 3-D phase less object detection follows a power iteration methodology and reduces the computational complexity to estimate the leading eigenvector of $\boldsymbol{\Lambda}$ using a low-pass filter $\boldsymbol{W}$. Bottom left: The unfolded network is subsequently trained, following setup in \cite{morales2022learning}, via backpropagation using paired inputs and outputs to optimize the filter $\boldsymbol{W}$. Right: Validation of the detection technique using experimental CDP acquired in the near zone ($\boldsymbol{A}_{1}$ in \eqref{eq:matrices})~\cite{ajerez}.}
		\vspace{-1em}
		\label{fig:unrolling3}
	\end{figure}

	For the 3-D object detection \cite{ajerez} based on phase-only information from CDP, a rapid estimation of the scene from CDP is carried out before detecting the target using its optical phase through cross-correlation analysis (template matching). For the estimation, \textit{filtered spectral initialization} approximates the phase of the 3-D object from the acquired CDP. This requires low-pass filtering of the leading eigenvector of matrix $\boldsymbol{\Lambda}=\frac{1}{\lvert \mathcal{I}_{0} \rvert} \sum_{i\in \mathcal{I}_{0}} \frac{\boldsymbol{a}_{k,i}\boldsymbol{a}_{k,i}^{H}}{\lVert \boldsymbol{a}_{k,i} \rVert_{2}^{2}}$. Here, unfolding (Figure~\ref{fig:unrolling3}) is employed to learn the low-pass filter \cite{morales2022learning}, where $\mathcal{I}_{0}$ is as in Section~\ref{subsec:withprior}. The low-pass filter which facilitates accurate estimation of the phase of the object is the key aspect of initialization.
	
	The filtered spectral initialization follows a power iteration method (see Figure \ref{fig:unrolling3}), wherein $\boldsymbol{z}^{(t+1)} = \boldsymbol{W}*(\boldsymbol{\Lambda}\boldsymbol{z}^{(t)})$ where $\boldsymbol{z}^{(0)}$ is chosen at random, $\boldsymbol{W}$ is the low-pass filter, and $*$ is the convolution operation. This product is then normalized. Iteratively applying $\boldsymbol{W}$ over the estimation of the image selects those low frequencies that sparsely represent the image in the Fourier domain. Since the filtering process is a convolution this selection is rapidly performed.
	
	In the next step, a circular harmonic filter $\boldsymbol{CHF}$ (in Fourier domain polar coordinates) is built based on a reference pattern of the scene \cite{ajerez}. This system of coordinates is preferred in order to make CHF invariant to rotations so that the filter detects the object regardless of any of its rotated version. The correlation matrix $\hat{\boldsymbol{C}}$ between $\boldsymbol{CHF}$ and Fourier transform of the approximated optical field $\hat{\boldsymbol{Z}}$ is then computed. The target is detected using a thresholding approach by building the decision matrix
	\begin{align}
		(\boldsymbol{R})_{u,v}=\left\{\begin{matrix}
			1, & \mathrm{if}\; \left|(\hat{\boldsymbol{C}})_{u,v}\right|\geq \varepsilon \cdot\max{(\hat{\boldsymbol{C}})} \\ 
			0, & \mathrm{otherwise}
		\end{matrix}\right.,
		\label{eq:decision}
	\end{align}
	where $\varepsilon\in(0,1]$ is the tolerance parameter. Figure~\ref{fig:unrolling3} shows the unfolded detection performance using experimental CDP in the near zone for a single snapshot. Here, the experimental data were acquired following the optical setup in \cite{katkovnik2017computational}.
	
	\begin{figure*}[t!]
		\centering
		\includegraphics[width=1\linewidth]{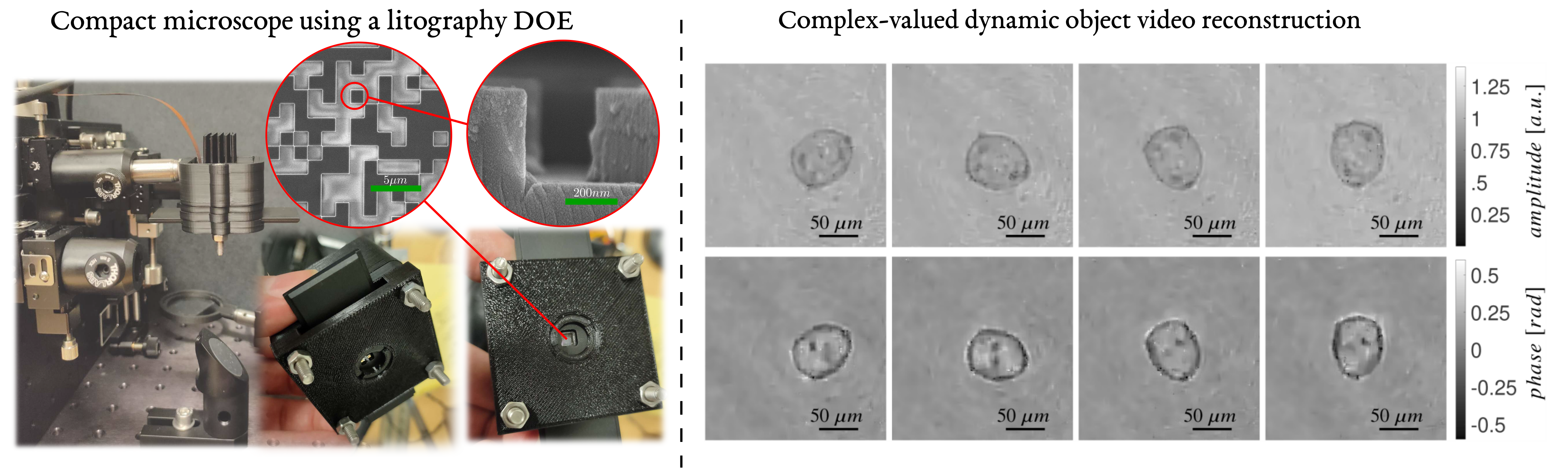}
		\vspace{-2em}
		\captionof{figure}{Left: Physical implementation of a compact microscope using DOE manufactured in Tampere's Photonics Laboratory with an electron beam lithography system on a fused silica glass and its fragments. The CDPs are acquired in the near zone ($\boldsymbol{A}_{1}$ in \eqref{eq:matrices}). When compared with a theoretically expected design, note that the fabricated DOE has imperfections arising from physical artefacts such as material imperfections and printing inaccuracies. The top and side views of scanning electron microscope images of the DOE are shown in the right half of the image and magnified further. Right: Frames from a complex-valued dynamic object video with amplitudes (in a.u. or airy units, top row) and phases (in radians or rad; bottom row) of a moving single-celled eukaryote reconstructed using sparsity-based filtering method. The video footage consisted of 287 frames through 10 seconds (\href{https://opticapublishing.figshare.com/articles/media/protozoa_mov/16743166}{see visualization}), of which frames 50, 100, 150, and 200 frames are reproduced (left to right) here.}
		\vspace{-1em}
		\label{fig:systems}
	\end{figure*}
	
	\subsection{Unfolded phase imaging}
	Phase imaging is of high interest in autonomous vehicles, detection of moving objects, and precision agriculture. Here, the 3-D shape of an object is recovered by solving the non-convex phase retrieval in a setup that records CDPs. Usually, acquisition of several snapshots from the scene is required. Recently, a single-shot technique was proposed in \cite{pinilla2020single}. It accurately estimates the optical phase $\varphi(x,y)$ across the spatial domain $(x,y)$ of the object following the pipeline of Figure~\ref{fig:unrolling3}. The estimated phase is then used to infer the 3-D shape by performing an unwrapping over $\varphi(x,y)$ to obtain $\varphi_{\textrm{unwrapp}}(x,y) = \varphi(x,y)+ 2k\pi,$ where $k$ is an integer that represents the projection period \cite{pinilla2020single}. The depth coordinate is based on the difference between the measured phase $\varphi_{\textrm{unwrapp}}$ and the phase value from a reference plane. This reference plane is obtained during the calibration of the acquisition system \cite{pinilla2020single}. Thus, the depth coordinate of the 3-D object is 
	\begin{align}
		\text{depth}(x,y) = c_{0} + c_{1}(\varphi_{\textrm{unwrapp}}(x,y) - \varphi_{0}(x,y)),
	\end{align}
	where $\varphi_{0}(x,y)$ is the reference phase and $\{c_{i}\}_{i=0}^1$ are tunable constants.
	
	\section{Enabling DOE Experimental Setups}
	
	The advantage of the DOE lies in its imaging capability to successfully recover the phase without additional optical elements (such as lenses) leading to even more compact imaging devices \cite{kocsis2021ssr}. The eschewing of the lens, which is the case in phase retrieval from CDP, makes the system not only light and cost-effective but also lens-aberration-free and with a larger field of view. Here, we summarize a few recent developments in experimental setups that employ DOE and are of interest to enable the use of unfolding-aided phase retrieval techniques.\\
	\textbf{Lensless super-resolution microscopy:} 
	In case of lensless super-resolution microscopy for middle zone CDP observations (Figure \ref{fig:systems} and Table \ref{tab:1}), the propagation operator $\mathcal{P}_{z}(\boldsymbol{w})$ at a distance $z$ is precisely modelled through convolution of a wavefront $\boldsymbol{w}$ with propagation kernel $\boldsymbol{K}_{z}$ as $\mathcal{P}_{z}(\boldsymbol{w})= \boldsymbol{F}^{-1} \left \{ \boldsymbol{K}_{z}\cdot\boldsymbol{F} \left \{ \boldsymbol{w} \right \} \right \}$, where
	\begin{equation}
		\label{eq:trfun}
		\boldsymbol{K}_{z}=\begin{cases} e^{i\frac{2\pi }{\lambda }z\sqrt{1-\lambda ^{2}\left ( f_{x}^{2} + f_{y}^{2}\right )}}, & f_{x}^{2} + f_{y}^{2}\leq \frac{1}{\lambda ^{2}},\\ 0, & \textrm{otherwise}.\\ \end{cases}
	\end{equation}
	Considering the spatial distribution of DOE, the observation model \cite{kocsis2021ssr} for the scene $\boldsymbol{x}$ and DOE $\boldsymbol{D}$ is $\boldsymbol{y} = \left \lvert \mathcal{P}_{z_{2}}\left(\boldsymbol{D}\cdot \mathcal{P}_{z_{1}}\left(\boldsymbol{x} \right) \right) \right \rvert^{2}$, where $z_{1}$ and $z_{2}$ are object-mask and mask-sensor distances, correspondingly (see Table \ref{tab:1}). In a phase retrieval with DOE setting, the object, mask, and sensor planes may have different sampling intervals $\Delta_{o}$, $\Delta_{m}$, and $\Delta_{s}$, respectively. If the resulting pattern $\boldsymbol{y}$ has computational pixels smaller than the physical sensor ($\Delta_{c} < \Delta_{s}$) and the reconstruction of the object is produced with $\Delta_{c}$, it leads to a subpixel resolution or super-resolution scenario. The ratio $r_{s}=\Delta_{s}/\Delta_{c}\geq1$ is the super-resolution factor. 
	
	With the objective of manufacturing the DOE, the matrix $\boldsymbol{D}$ is modelled as $\mathbf{D}_{s,u}=\text{exp}(j\phi_{s,u})$, for $u,s=1,\dots,n$, where $\phi_{s,u}$ takes value from a uniform random variable $d$ with two possible equi-probable coding elements $0$ and $2.62\texttt{rad}$, i.e, $d=\{\exp(j0),\exp(2.62j\texttt{rad})\}$. These values of $d$ are chosen because they lead to the best performance in terms of reconstruction quality \cite{kocsis2021ssr} (recall $\mathbb{E}[d]\not =0$). Figure~\ref{fig:systems} shows the resulting DOE (manufactured in Tampere's Photonics Laboratory) with the pixel size of $\Delta_{m}=1.73$ $\mu$m. This size is half of the sensor and, hence, a better fit for the super-resolution factor $r_s$ in powers of 2. The compact microscope in Figure~\ref{fig:systems} enables full wavefront reconstruction of dynamic scenes. Contrary to the existing phase retrieval setups, this unique system is capable of computational super-resolved video reconstruction via phase retrieval with a high frame rate limited only by the camera. 
	
	To reconstruct the signal of interest as reported in Figure~\ref{fig:systems} (bottom right), \cite{kocsis2021ssr} studied the impact of a special multiplicative calibration term, which allows building a good propagation model to compensate errors of theoretical models. This corrective term was shown to improve the imaging quality and resolution. However, it is constant across iterations. Therefore, it may benefit from a reconditioning unfolded method, wherein the correction is varied on per iteration basis.
	
	\begin{figure}[t!]
		\centering
		\includegraphics[width=0.9\linewidth]{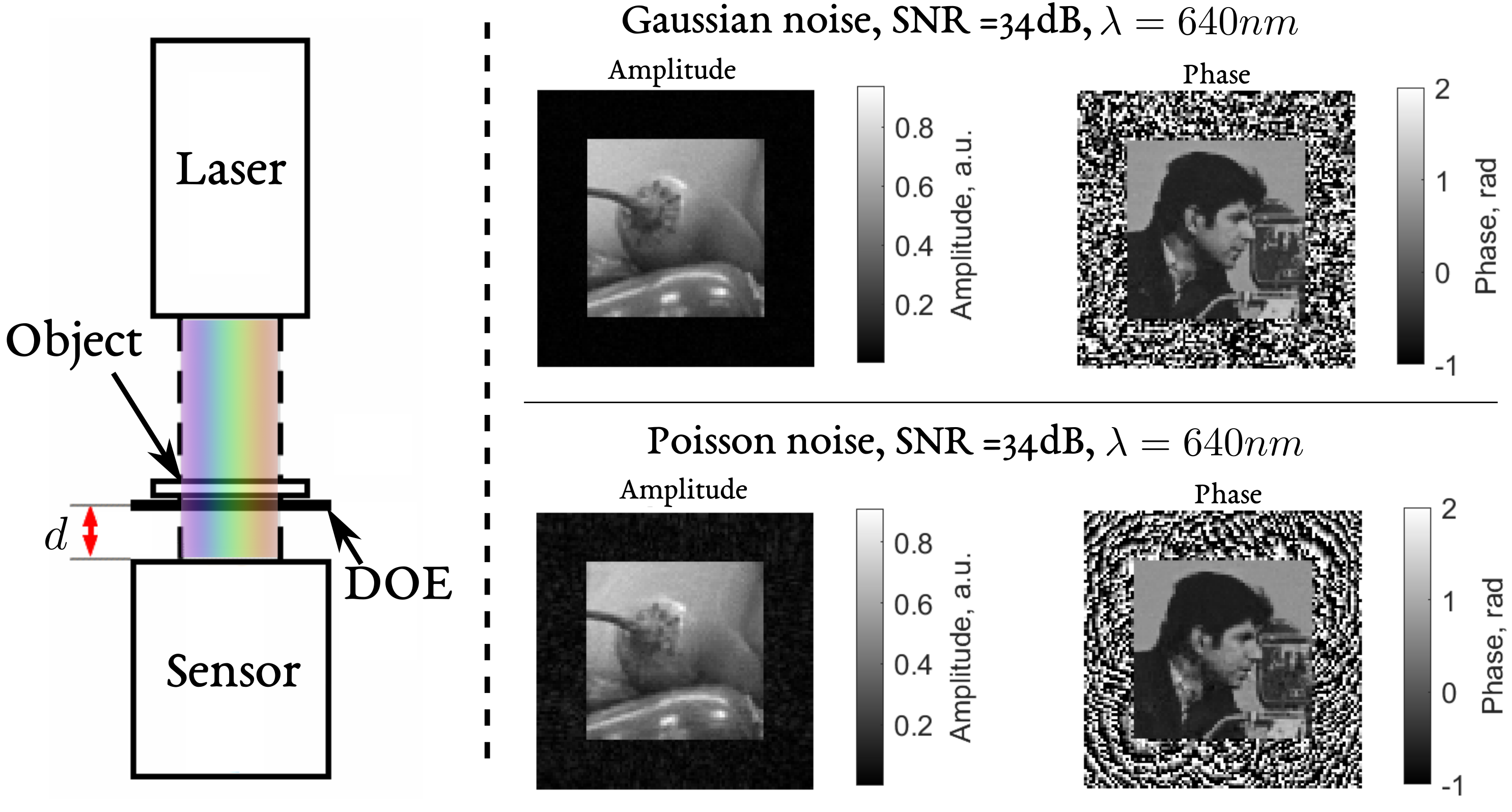}
		\vspace{-1em}
		\captionof{figure}{Left: Hyperspectral optical setup: ``Laser" is a broadband coherent light source, ``DOE" is modelled as a spatial light modulator, and ``Sensor" is a registration camera. Right: Reconstructed phase and amplitude from CDP in near zone ($\boldsymbol{A}_{1}$ in \eqref{eq:matrices}) for the wavelength $\lambda=640$ nm in the presence of Gaussian (top right) and Poisson noise (bottom right) with $SNR=34$ dB in each case. The images are shown in square frames to emphasize that the size and location of the object support are assumed unknown and reconstructed automatically. The support of true image is used only for computation of observations produced for the zero-padded object. For the amplitudes, these frames have nearly zero values, while the phase estimates take random values in the frame area as the phase cannot be defined when amplitudes vanish.} \vspace{-0.5em}
		\label{fig:complexHS}
	\end{figure}
	
	\begin{figure}[ht]
		\centering
		\includegraphics[width=0.9\linewidth]{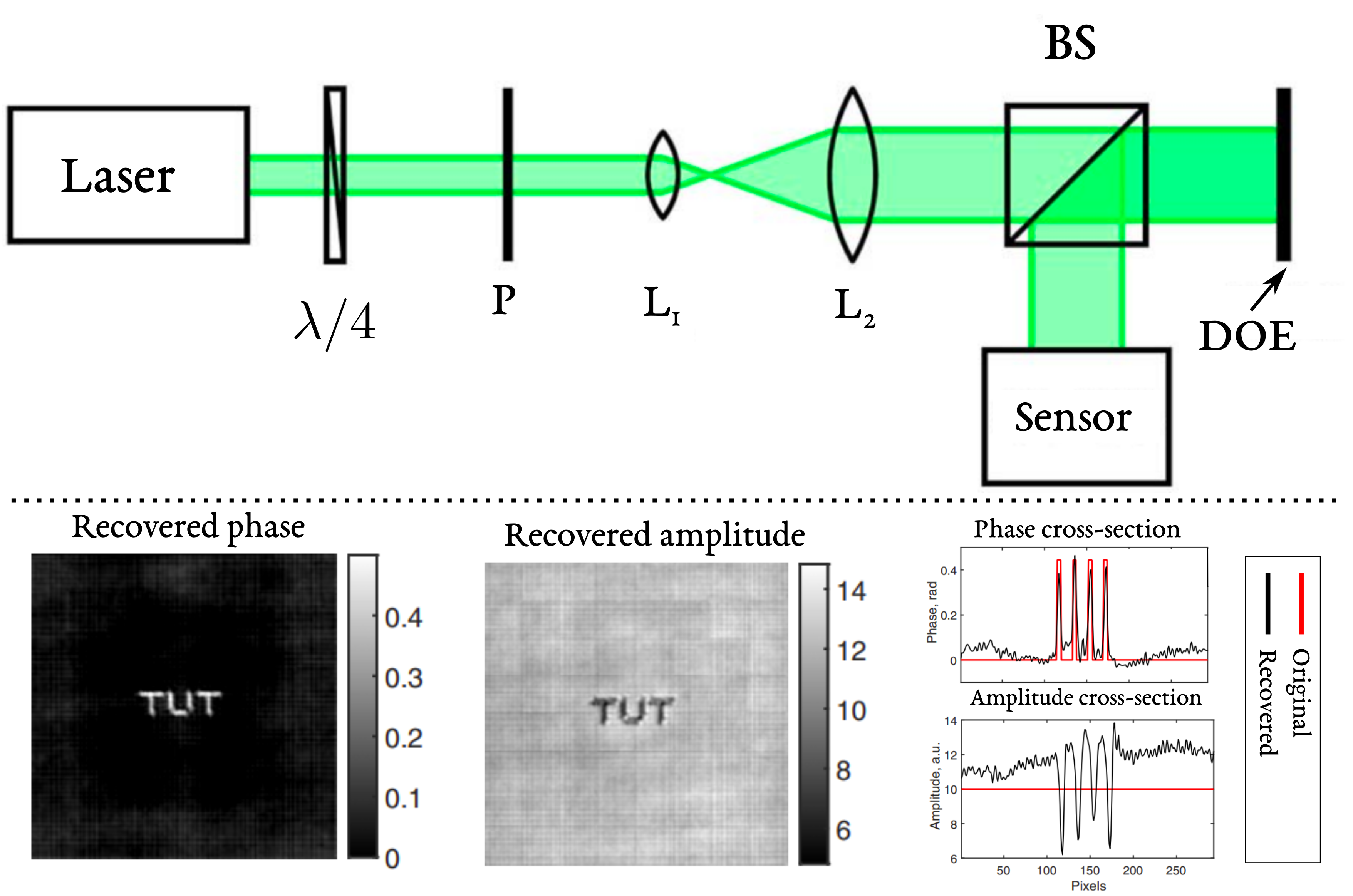}
		\vspace{-0.5em}
		\caption{\small{Top: Blind deconvolution super-resolution phase retrieval optical setup comprising $\lambda/4$ retardation plate, polarizer ($P$), lenses ($L_{1}$ and $L_{2}$), beamsplitter (BS), DOE physically implemented using a spatial light modulator (SLM), sensor, and registration camera. Bottom: Recovered amplitude and phase of the object ``TUT" from CDP at near zone ($\boldsymbol{A}_{1}$ in \eqref{eq:matrices}) with pixel-wise resolution factor of $4$ \cite{katkovnik2017computational}. The SLM was configured to work as a phase-only modulator because the device is not very accurate in modulating both magnitudes and phases \cite{holoeye}. }}
		\vspace{-1em}
		\label{fig:phasu}
	\end{figure}
	
	\textbf{Hyperspectral complex-domain imaging:} 
	This is a comparatively new development that deals with a phase delay of a coherent light in transparent or reflective objects \cite{katkovnik2021admm}. The hyperspectral broadband phase imaging is more informative than the monochromatic technique (see near zone setup and results in Figure \ref{fig:complexHS}). Conventionally, for the processing of hyperspectral images, 2-D spectral narrow-band images are stacked together and represented as 3-D cubes with two spatial coordinates $(x,y)$ and a third longitudinal spectral coordinate. In hyperspectral phase imaging, for a given wavelength $\lambda$, data $\boldsymbol{x}_{\lambda}\in \mathbb{C}^{n}$ in these 3-D cubes are complex-valued with spatially and spectrally varying amplitudes and phases. Denote the set of wavelengths under study by $\mathcal{L}$. From \eqref{eq:matrices}, the measurement matrix for diffraction zone $k$ and wavelength $\lambda$ is $\boldsymbol{A}_{k,\lambda}$. Then, the noisy phaseless measurements are \cite{katkovnik2021admm}
	\begin{align}
		\boldsymbol{y}_{k} = \sum_{\lambda\in \mathcal{L}}\lvert \boldsymbol{A}_{k,\lambda} \boldsymbol{x}_{\lambda} \rvert^{2} + \boldsymbol{\epsilon}.
	\end{align}
	The reconstruction of $\boldsymbol{x}_{\lambda}$ here is more ill-conditioned than the classical phase retrieval problem because the acquired data are summed across wavelengths of phaseless measurements. This sum does not appear in the classical phase retrieval, which employs single wavelength. This also makes phase image processing more complex than the hyperspectral intensity imaging, where the corresponding 3-D cubes are real-valued.
	
	The model-based algorithms require significant amount of measurements to efficiently recover complex-valued $\boldsymbol{x}_{\lambda}$ \cite{katkovnik2021admm}. As suggested by the previously mentioned numerical results, the preconditioning unfolding is useful in the hyperspectral complex-domain phase retrieval problem because it facilitates reducing the condition number, admitting fewer data samples, and converging quickly.
	
	\textbf{Blind deconvolution super-resolution phase retrieval:} 
	In certain CDP applications, the combination of blind deconvolution, super-resolution, and phase retrieval naturally manifests. While this is a severely ill-posed problem, it has been shown \cite{pinilla2021nonconvex} that image-of-interest could be estimated in polynomial-time. The approach relies on previous results that established the DOE design to achieve high quality images \cite{jbacca} and partially analyzed the combined problem by solving the super-resolution phase retrieval problem \cite{katkovnik2017computational} from CDP (Figure \ref{fig:phasu}). These designs are obtained by exploiting the model of the physical setup using ML methods where the DOE is modelled as a layer of a neural network (data-driven model or unrolled) that is trained to act as an estimator of the true image \cite{rostami2021power}. This data-driven design has shown an outstanding imaging quality using a single snapshot as well as robustness against noise.
	
	\section{Summary and future outlook}
	The plethora of phase retrieval algorithms and their relative advantages make it difficult to conclude which of them will always be the best. The only foregone conclusion has been that the computational imaging community needs to continue basing these algorithms on models devoid of several practical measurement errors. In this context, deep unfolding methods are gaining salience in optical phase retrieval problems to offer interpretability, state-of-the-art performance, and high computational efficiency. By learning priors from the trained data, these empirically designed architectures are exploited to build globally converged learning networks for a specific phase retrieval setting. A particularly desired outcome of this development is the improvement in the design of DOEs that satisfies theoretical conditions for the uniqueness of signal recovery. 
	
	We provided examples of several recent novel setups which indicate that unfolding has the potential to enable a new generation of optical devices as a consequence of co-designing optics and recovery algorithms. Several applications of lensless imaging, synthetic apertures, and highly ill-posed variants of phase retrieval appear within reach in the near future using this technique, including in areas such as optical metrology, digital holography, and medicine.
	
	\bibliographystyle{IEEEtran}
	\bibliography{main}
	
\end{document}